\documentclass[10pt,journal,compsoc]{IEEEtran}

\ifCLASSOPTIONcompsoc
  \usepackage[nocompress]{cite}
\else
  \usepackage{cite}
\fi

\ifCLASSINFOpdf
 
\else
 
\fi

\usepackage{lipsum}
\usepackage{ragged2e}
\justifying
\usepackage{framed}
\usepackage{threeparttable}
\usepackage{tabularx}
\usepackage{diagbox}
\usepackage{fancyhdr}
\usepackage[normalem]{ulem}
\usepackage{hyperref}
\usepackage{amsmath}
\usepackage{booktabs} 
\usepackage{upgreek}
\usepackage{graphicx}
\usepackage{verbatim}
\usepackage{multirow}
\usepackage{caption}
\usepackage{array}
\usepackage{subfigure}
\usepackage{float}
\usepackage{CJK}
\usepackage{setspace}
\usepackage[font=footnotesize]{caption}
\usepackage{indentfirst}
\usepackage{dblfloatfix} 
\usepackage{multirow}
\usepackage[noend]{algpseudocode}
\usepackage{algorithmicx,algorithm}
\usepackage{xcolor}
\graphicspath{{./Fig/}}
\newcommand*\circled[1]
            {\raisebox{.5pt}{\textcircled{\raisebox{-.9pt} {#1}}}}

  \DeclareSymbolFont{numbers}{T1}{ppl}{m}{n}
  \SetSymbolFont{numbers}{bold}{T1}{ppl}{bx}{n}
  \DeclareMathSymbol{0}\mathalpha{numbers}{"30}
  \DeclareMathSymbol{1}\mathalpha{numbers}{"31}
  \DeclareMathSymbol{2}\mathalpha{numbers}{"32}
  \DeclareMathSymbol{3}\mathalpha{numbers}{"33}
  \DeclareMathSymbol{4}\mathalpha{numbers}{"34}
  \DeclareMathSymbol{5}\mathalpha{numbers}{"35}
  \DeclareMathSymbol{6}\mathalpha{numbers}{"36}
  \DeclareMathSymbol{7}\mathalpha{numbers}{"37}
  \DeclareMathSymbol{8}\mathalpha{numbers}{"38}
  \DeclareMathSymbol{9}\mathalpha{numbers}{"39}

\begin{document}

\title{\Huge{Neural-PIM: Efficient Processing-In-Memory with Neural Approximation of Peripherals}}

\author{Weidong Cao, Yilong Zhao, Adith Boloor, Yinhe Han, Xuan Zhang, and Li Jiang
\IEEEcompsocitemizethanks{\IEEEcompsocthanksitem W. Cao, and X. Zhang are with the Department of Electrical and Systems Engineering; A. Boloor is with the Department of Computer Science and Engineering, Washington University in St. Louis, MO, 63130 USA.
\IEEEcompsocthanksitem Y. Zhao, and L. Jiang are with the Department of Computer Science and Engineering, Shanghai Jiao Tong University, Shanghai, 200240 CHN.
\IEEEcompsocthanksitem Y. Han is with the Institute of Computing Technology, Chinese Academy of Sciences, Beijing, 100190 CHN.

}
\thanks{Manuscript received 11 Sept. 2020; revised 30 Mar. 2021; accepted 10 Oct. 2021, IEEE Transactions on Computers.}  
\thanks{(Corresponding authors: $\{$xuan.zhang@wustl.edu, jiangli@cs.sjtu.edu.cn$\}$).}
}

\IEEEtitleabstractindextext{%
\begin{abstract}
\justifying{Processing-in-memory (PIM) architectures have demonstrated great potential in accelerating numerous deep learning tasks.
Particularly, resistive random-access memory (RRAM) devices provide a promising hardware substrate to build PIM accelerators due to their abilities to realize efficient in-situ vector-matrix multiplications (VMMs).
However, existing PIM accelerators suffer from frequent and energy-intensive analog-to-digital (A/D) conversions,
severely limiting their performance.
This paper presents a new PIM architecture to efficiently accelerate deep learning tasks by minimizing the required A/D conversions with analog accumulation and neural approximated peripheral circuits.
We first characterize the different dataflows employed by existing PIM accelerators, based on which a new dataflow is proposed to remarkably reduce the required A/D conversions for VMMs by extending shift and add (S+A) operations into the analog domain before the final quantizations.
We then leverage a neural approximation method to design both analog accumulation circuits (S+A) and quantization circuits (ADCs) with RRAM crossbar arrays in a highly-efficient manner.
Finally, we apply them to build a RRAM-based PIM accelerator (i.e., \textbf{Neural-PIM}) upon the proposed analog dataflow and evaluate its system-level performance.
Evaluations on different benchmarks demonstrate that Neural-PIM can improve energy efficiency by $5.36\times$ ($1.73\times$) and speed up throughput by $3.43\times$ ($1.59\times$) without losing accuracy, compared to the state-of-the-art RRAM-based PIM accelerators, i.e., ISAAC~\cite{isaac} (CASCADE~\cite{cascade}).}
\end{abstract}
}

\maketitle

\IEEEdisplaynontitleabstractindextext

\IEEEpeerreviewmaketitle

\section{Introduction}
\label{sec:intro}
\IEEEPARstart{D}{eep} neural networks (DNNs) have been powering a broad range of applications~\cite{zhu2018cmos,feng_1,NLP,alexnet,overfeat}, including natural language processing, image classification, and object recognition.
The demand to achieve high accuracy for increasingly computational tasks leads to ever-growing model sizes of modern DNNs~\cite{NIPS2019_fix,ECCV_18}.
Processing such huge DNNs~\cite{NLP,alexnet, overfeat,NIPS2019_fix,ECCV_18} in systems with conventional Von Neumann architectures~\cite{an_1,an_2,chen_1} incurs enormous energy consumption and significant execution latency due to the vast data movement between the separate memory and computing elements.
To tackle this challenge, processing-in-memory (PIM) is introduced as a promising paradigm by co-locating
compute and memory.
Various PIM architectures have been proposed to accelerate DNNs~\cite{promise,compute_caches,drisa,isaac,prime,pipelayer,cascade,sparse_ReRAM} in the past years.
Particularly, emerging non-volatile memory devices, e.g., resistive random access memory (RRAM) device, 
have been extensively studied to design PIM accelerators owing to 
their inherent advantage in realizing highly efficient in-situ vector-matrix multiplication (VMM) with crossbar arrays.

Examples of such accelerators~\cite{isaac,prime,pipelayer,cascade,sparse_ReRAM} have demonstrated significant improvements in energy efficiency and throughput on the order of $\sim$1000$\times$ compared to CPU and GPU platforms.
Despite of their promise, RRAM-based PIM accelerators are still in the early stage of development with many open challenges.
One primary concern is that they rely on costly peripheral circuits, e.g., digital-to-analog converters (DACs) and analog-to-digital converters (ADCs), to bridge analog VMM and digital storage. 
Given the typical size of a crossbar array ranging from $64\times 64$ to $256\times 256$ and the precision of a RRAM cell varying from 1-bit to 6-bit~\cite{isaac,prime, pipelayer, cascade, sparse_ReRAM,enabling_sci}, analog partial sums on bitlines (BLs) could have extremely fine voltage/current levels.
Such partial sums demand high-resolution ADCs for quantization which however dominate the energy consumption and the silicon area of RRAM-based PIM accelerators.
For example, 58$\%$ of system energy is consumed by 8-bit ADCs in ISAAC~\cite{isaac}, and 98$\%$ of silicon area is occupied by 8-bit ADCs in a scientific computing accelerator~\cite{enabling_sci}.

Prior work has reported a number of various techniques to alleviate the heavy burden of ADCs in RRAM-based PIM accelerators.
CNNWire~\cite{CNNWire} uses the Winograd algorithm to reduce the number of required analog VMM operations, thereby decreasing the associated energy consumed by analog-to-digital (A/D) conversions.
Yet, without optimizing the peripherals of crossbar arrays holistically, this method still endures huge overheads caused by its peripherals such as ADCs.
PRIME~\cite{prime} and PipeLayer~\cite{pipelayer} adopt single-bit sense amplifiers and integrate-and-fire (IF) neurons to replace traditional high-resolution ADCs.
However, these 1-bit quantizers take up to $2^n$ clock cycles to produce an $n$-bit BL output, leading to significant processing latency.  
CASCADE~\cite{cascade} leverages RRAM buffer arrays to temporally store the analog partial sums and accumulate them before quantization.
This method can effectively reduce the required A/D conversions.
However, buffering high-precision (6-bit) analog partial sums into RRAM cells demands non-trivial programming energy and suffers from severe device variations~\cite{rram_approxi,li2015variation}, degenerating the inference accuracy.
Without losing accuracy and throughput, further energy and area improvements in peripherals, are of paramount importance in building energy-efficient and high-performance RRAM-based PIM accelerators.

In this paper, we aim to fundamentally address the overheads of RRAM-based PIM accelerators caused by the expensive A/D conversions. 
Towards the goal, we develop an analytical framework to characterize different accumulation strategies employed by existing RRAM-based PIM accelerators.
In this way, we reveal opportunities to further improve their performance by extending the accumulation into the analog domain to minimize the need for explicit A/D conversions.
With this key insight, we propose \textbf{Neural-PIM}--a novel accelerator architecture with \textbf{neural} approximated peripheral circuits to improve both energy efficiency and throughput for RRAM-based \textbf{PIM} acceleration.
Key innovations and contributions in the paper are listed below:
\begin{itemize}
    \item \textcolor{black}{We classify different accumulation schemes used by RRAM-based PIM accelerators and build a unified analytical framework to compare the performances of associated dataflows.
    The study shows that our proposed analog dataflow which fully extends the accumulation of partial sums into the analog domain can minimize the required A/D conversions.
    }
    \item We propose a novel neural approximation method to design peripheral circuits (termed NeuralPeriph), enabling the proposed extended analog dataflow. 
    NeuralPeriph circuits are synthesized using RRAM crossbar arrays and CMOS inverters, hence, are both energy- and area-efficient.
    \item We build a Neural-PIM accelerator upon the proposed analog dataflow with NeuralPeriph circuits.
    Particularly, we present the detailed system design and thoroughly discuss the inference accuracy of the proposed accelerator. 
    \item  We comprehensively investigate its system-level performance.
    Compared to the state-of-the-art baselines, i.e., ISAAC~\cite{isaac} (CASCADE~\cite{cascade}),
    Neural-PIM shows an average improvement of $5.36\times$ ($1.73\times$) in energy efficiency, and $3.43\times$ ($1.59\times$) in throughput across different DNN benchmarks without losing accuracy.
     \end{itemize}

The rest of the paper is organized as following:
Section~\ref{sec:bg} provides the background of the work.
Section~\ref{sec:charac} introduces the characterization framework.
NeuralPeriph circuits are presented in Section~\ref{sec:nnperi}.
Neural-PIM accelerator is elaborated in Section~\ref{sec:architecture}.
Finally, we present the simulation methodology in Section~\ref{sec:ex_metho} and demonstrate the evaluation results in Section~\ref{sec:experiment} before concluding the paper in Section~\ref{sec:con}.
\section{Preliminary Knowledge}
\label{sec:bg}

In this section, we first review typical DNN structures and associated operations. 
We then briefly show basic architectures and components of RRAM-based PIM accelerators.
Finally, we introduce the concept of
neural approximators.

\subsection{Neural Network Workloads}
\label{sec:workload}

Convolutional neural networks (CNNs) and recurrent neural networks (RNNs) are two popular DNN workloads.

\begin{figure}[!t]
 \centering
    \includegraphics[width=1.0\linewidth]{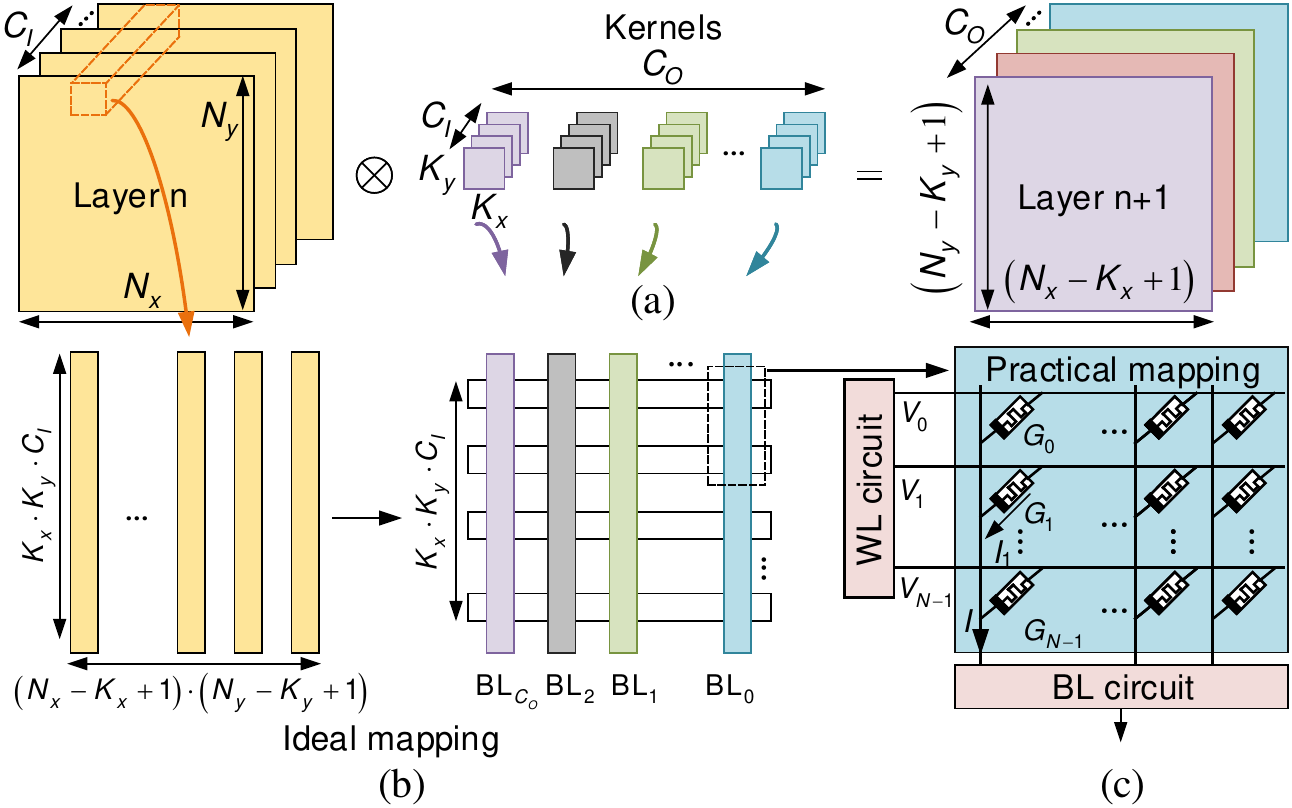}
    \caption{Mapping illustration. (a), An example of a CONV layer. (b), Ideally mapping a CONV layer upto crossbar arrays. (c), Practical mapping and simplified illustration of peripherals at the array level.}
    \label{fig:rram_mapping}
\end{figure}

\textbf{{CNN models.}}
A typical CNN consists of many cascaded computational layers, such as convolutional (CONV) layers, pooling layers (POOL), and fully connected (FC) layers. 
CONV and FC represent the most computation-intensive and memory-intensive layers as they involve large amounts of multiplication and accumulation operations.
Fig.~\ref{fig:rram_mapping}(a) illustrates an exemplary CONV layer in a CNN. 
The input feature map of layer $n$ is a 3-D tensor defined with a size of $N_x \times N_y \times C_{I}$. 
The kernels are a 4-D tensor with a size of $K_x \times K_y \times C_{I} \times C_{O}$. 
Here, $C_{I}$ and $C_{O}$ are the number of input channels and output channels, respectively.
To get the output feature map of layer $n+1$, the kernel window shifts right or down to convolve the input feature map.
An output point $f^{O}_{x,y,z}$ at the position of $(x,y,z)$ is obtained as follow: $f^{O}_{{x,y,z}}= \sigma(\sum^{C_{I}}_{k=1}\sum^{K_{x}}_{i=1}\sum^{K_{y}}_{j=1}f^{I}_{x,y}(i,j,k)  \cdot w_z(i,j,k))$.
Here, $f^{I}_{x,y}(i,j,k)$ is an input; $w_z(i,j,k)$ is the weight value of the $z^{th}$ kernel; $\sigma(\cdot)$ is an activation function; $(x,y,k)$ and $(i,j,k)$ correspond to the positions of input and weight.
An FC layer can be viewed as a special case of a CONV layer when $N_x = K_x$ and $N_y = K_y$.
The corresponding output feature map is thus sized into $1\times 1\times C_I\times C_O$.

\textbf{RNN models.} Long-short term memory (LSTM)~\cite{lstm} and gated recurrent unit (GRU)~\cite{gru} are two representative RNNs.
The underlying computations of RNNs are similar to the FC layer in a CNN, which are expressed as the following equations by taking LSTM as an example:
\begin{equation}
\begin{split}
& \text{Input gate:}~ i_t =\sigma(W^i\cdot x_t+U^i\cdot h_{t-1});  \\
& \text{Forget gate:}~ f_t =\sigma(W^f\cdot x_t+U^f\cdot h_{t-1});   \\
& \text{Output gate:}~ o_t =\sigma(W^o\cdot x_t+U^o\cdot h_{t-1}); \\
&  \text{Cell state:}~ \tilde c_t =\tanh(W^c\cdot x_t+U^c\cdot h_{t-1});  \\
&  \text{Memory cell:}~ c_t =f_t\odot c_{t-1}+i_t\odot \tilde c_t; \\
& \text{Hidden state:}~ h_t =o_t\odot \tanh(c_t).
\end{split}
\label{eqn:lstm}
\end{equation}
Here, $W^{k}$ and $U^{k}$, $k=\{i,f,o,c\}$ are the learned weights for input vector $x_t$ and hidden states $h_{t-1}$ of different gates respectively.
$\odot$ denotes element-wise multiplication.
Both $\sigma$ and $\tanh$ are the activation functions.
The first four rows in Eq.~\eqref{eqn:lstm} involve considerable computation-intensive VMM operations.
As the time step $t$ increases, the computations and memory accesses increase linearly.

\textbf{DNN quantization.}
DNN models are usually quantized into low-precision, e.g., 16-bit fixed-point numbers~\cite{haq,bit_fusion}, to reduce the model size and the complexity of hardware.
The quantization enables better performances with negligible accuracy loss. 
Recent work~\cite{haq,quan_1,quan_2} has shown that lower-precision ($\leq$ 8-bit) quantizations can achieve higher energy efficiency while keeping the same accuracy as the 16-bit quantization for the PIM accelerators.

\subsection{RRAM-Based PIM Architecture}
\label{sec:tra_peri}

A number of RRAM-based PIM accelerators have been reported in prior work~\cite{isaac,prime,pipelayer,cascade, sparse_ReRAM, enabling_sci}.
They have various architectures and execution manners, but possess similar devices, building blocks, and peripheral circuitry modules (including wordline circuitry and bitline circuitry) at the array level as shown in Fig.~\ref{fig:rram_mapping}(c).

\textbf{{RRAM device.}}
A RRAM device is a passive element which stores information with its conductance.
Taking advantage of its small size and excellent scalability, RRAM is usually organized into a dense crossbar architecture, serving as either a storage array or a computing engine.

\textbf{{RRAM crossbar array.}}
A RRAM crossbar array allows to parallelly perform massive VMMs in the analog domain.
Fig.~\ref{fig:rram_mapping}(b) presents an example of mapping a CONV layer into RRAM crossbar arrays for in-situ VMM. 
The $K_x\times K_y\times C_I$ weights of each kernel can be stored into the same number of RRAM cells as conductances in a column.
The $C_O$ kernels are then mapped into $C_O$ columns.
Measured data from RRAM chips show that both RRAM precision (1$\sim$3-bit) and crossbar array size ($\leq 256\times256$) are limited~\cite{rram_pre}.
A practical mapping method is to split each high-precision weight into several RRAM cells of adjacent columns and store a large-size kernel across several crossbar arrays~\cite{isaac,prime,pipelayer,cascade,sparse_ReRAM, enabling_sci}.
By applying read voltage to the wordlines (WLs) of a crossbar array,
Ohm's law dictates that the corresponding current contributed by the RRAM cell to the BL is the product of the input voltage and the conductance.
Then, according to Kirchhoff's law, the current from each cell aggregates along the BL to complete the computation of a dot-product.

\textbf{{Wordline circuitry.}}
DACs are the WL drivers to convert digital inputs into analog voltages.
Employing high-resolution DACs as input drivers improves latency, but requires additional hardware resources and leads to high energy consumption.
Bit slicing technique serves an efficient input streaming strategy, where bit-slices of a high-precision input are serially fed to the WLs using a low-resolution DAC.
For example, ISAAC~\cite{isaac} and CASCADE~\cite{cascade} adopt a 1-bit DAC to stream a 16-bit input with 16 cycles.

\textbf{{Bitline circuitry.}}
Sense amplifiers (SAs)~\cite{prime}, integrate-and-fire (IF) neurons~\cite{pipelayer}, or ADCs~\cite{isaac} serve as BL quantizers to convert analog partial sums into digital bits. 
For example, PRIME~\cite{prime} and PipeLayer~\cite{pipelayer} adopt 1-bit SAs and IF neurons to successively perform $2^n$ conversions to produce an $n$-bit output.
Despite being more energy-efficient, these 1-bit quantizers have significant conversion latency.
On the other hand, conventional ADCs can obtain multi-bit quantization in one clock period.
ISAAC~\cite{isaac} uses 8-bit ADCs for quantization.
However, high-resolution ADCs lead to higher power and area overheads compared to SAs and IF neurons.
After quantization, shift and add circuits are used to tally up the partial sums across different BLs and different computation cycles to obtain the final dot-product.

\begin{figure}[!t]
    \centering
    \includegraphics[width=\linewidth]{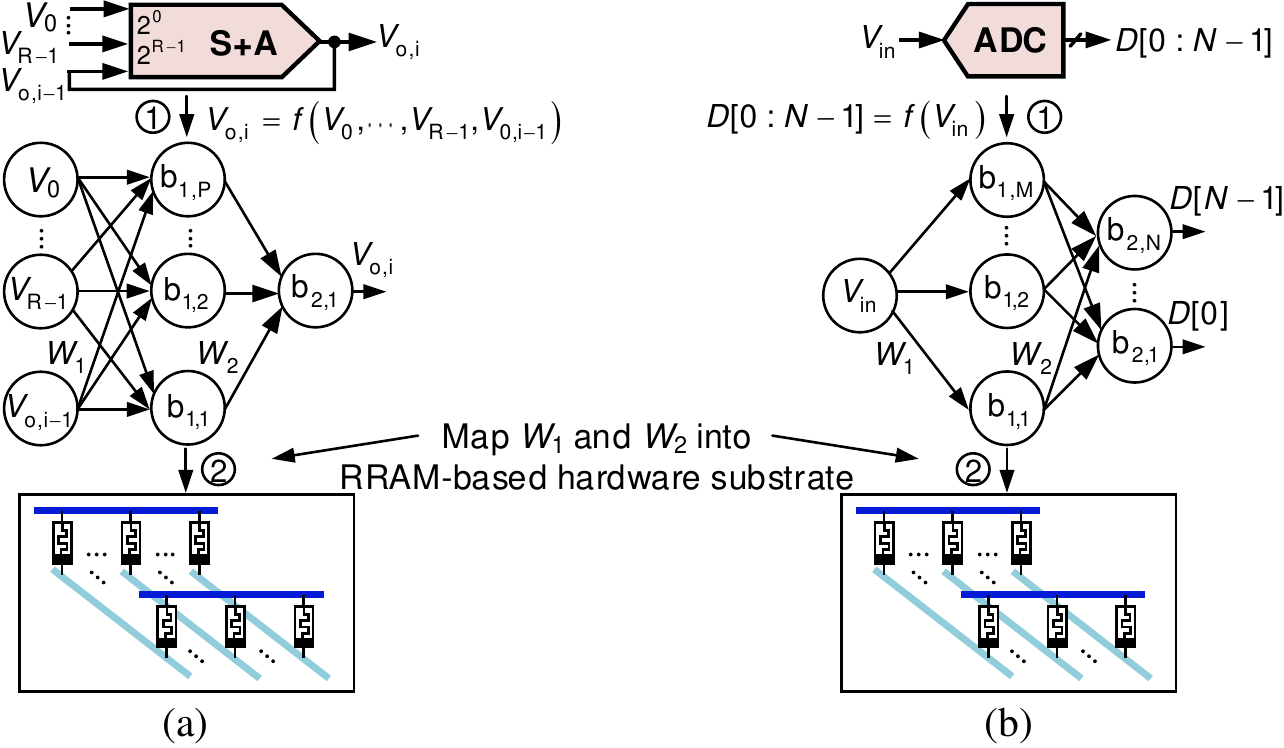}
    \caption{Conceptual illustration to design neural approximated peripheral circuits in RRAM-based PIM accelerator. (a), Neural approximation for S+A circuits. (b), Neural approximation for ADCs.}
    \label{fig:peripheralcircuit}

\end{figure}

\begin{figure*}[!t]
 \centering
    \includegraphics[width=1.0\linewidth]{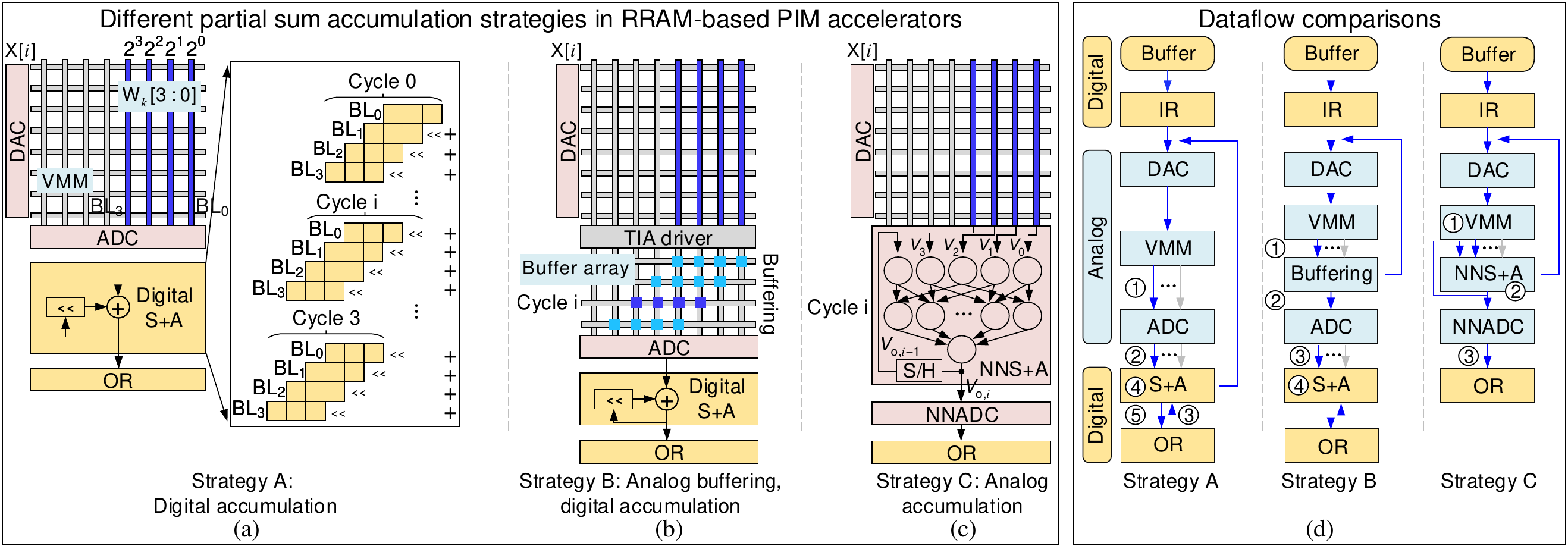}
    \caption{Different accumulation strategies in RRAM-based PIM accelerator~\cite{prime,isaac,pipelayer,cascade,sparse_ReRAM, enabling_sci} and their dataflows. Here, we use 4-bit dot-product as an example. (a), Strategy A with fully digital accumulation. (b), Strategy B with analog buffering and digital accumulation. (c), Proposed Strategy C with fully analog accumulation. (d), Dataflow comparisons between different dataflows. IR/OR are the input/output register.}
    \label{fig:accu_compa}

\end{figure*}

\subsection{Neural Approximator}
\label{sec:Neu_appro}

A neural approximator refers to a specialized hardware that approximates a general function $\vec{y}=f(\vec{x})$ in its circuit implementation.
The underlying idea stems from the universal approximation theorem~\cite{u_approx} that states an arbitrary decision region can be well-approximated via training a feedforward neural network (NN) with one hidden layer and any continuous nonlinear activation function (NAF).
In other words, with $n_1$ inputs and $n_2$ outputs, a neural approximator built upon a three-layer NN is capable of approximating any
$n_1$-input-$n_2$-output function~\cite{era-lstm}.
Prior work~\cite{rram_approxi,cao2019neuadc,cao2019neuadc_C,pipelineadc,pipelineadc_J} has leveraged RRAM crossbar arrays and analog neurons to design high-performance neural approximators for diverse applications.
For example, neural approximators can accelerate various computational tasks with high accuracy ($<$~$10^{-6}$ mean square errors) and superior energy efficiency (24.59$\sim$567.98 GFLOPS/W)~\cite{rram_approxi}.
They can also approximate diverse quantization functions with remarkable performance~\cite{cao2019neuadc,pipelineadc}.

Inspired by prior success of applying neural approximators to many computation and quantization tasks, we explore a new direction of neural approximators in the design of peripheral circuits, e.g., shift and add (S+A) circuits and converters (ADCs), to improve the performance of RRAM-based PIM accelerators.
Fig.~\ref{fig:peripheralcircuit} conceptually shows the idea to design the neural approximated peripherals.
A three-layer NN is first trained to approximate the mathematical function of each peripheral circuit, i.e., a conventional shift and add function for an S+A circuit and a quantization function for an ADC (\circled{1}).
The well-trained NN is then mapped into RRAM-based hardware substrate (\circled{2}).
We elaborate the design of these neural approximated circuits in Section~\ref{sec:nnperi}.

\section{Characterizing PIM Dataflows}
\label{sec:charac}

In this section, we examine the partial sum accumulation schemes employed by existing RRAM-based PIM accelerators and their associated dataflows.
In this way, we identify a new dataflow that can perform fully-analog accumulation. 
We then present a unified analytical framework to study the first-order performance associated with each dataflow.
This modeling framework allows us to capture the impact of the main design parameters on the system performance (i.e., accuracy, energy, and latency) and 
quantify the key benefits of the proposed analog accumulation dataflow.

\subsection{Classifying Accumulation Schemes}
\label{sec:Per}
We find the accumulation schemes of partial sums significantly affect the energy efficiency of RRAM-based PIM accelerators.
Fig.~\ref{fig:accu_compa}(a)-(c) show the different accumulation strategies.
Fig.~\ref{fig:accu_compa}(d) compares the corresponding dataflows.

Strategy A (Fig.~\ref{fig:accu_compa}(a)) conducts accumulation after quantizing BL analog partial sums.
Prior work, e.g., ISAAC~\cite{isaac}, PRIME~\cite{prime}, and PipeLayer~\cite{pipelayer}, adopts this strategy.
\textcolor{black}{Its dataflow (Strategy A in Fig.~\ref{fig:accu_compa}(d)) follows below steps for each input bit-slice vector $X[i]$:
\circled{1} Transfer BL currents of a VMM computing array to voltages;
\circled{2} Converter the analog voltages to digital partial sums;
\circled{3} Read out the temporary sums stored in a local OR;
\circled{4} Accumulate the partial sums with S+A circuits;
\circled{5} Write back the updated sum to the OR.}
Assume using 1-bit RRAM cells to store an 8-bit weight, the partial-sum accumulation needs $8\times 8=64$ A/D conversions and data movements in and out of registers to obtain an 8-bit dot-product, which significantly worsens the energy efficiency.
Strategy B (Fig.~\ref{fig:accu_compa}(b)) buffers analog partial sums from all input cycles before the quantization, capturing the scheme adopted by CASCADE~\cite{cascade}.
\textcolor{black}{The dataflow (Strategy B in Fig.~\ref{fig:accu_compa}(d)) follows below sequences for every input bit-slice vector $X[i]$: \circled{1} Transfer BL currents of a VMM computing array to voltages using trans-impedance amplifiers (TIAs);
\circled{2} Align the voltages as inputs to write RRAM in buffer arrays to store the partial sums.
After the analog partial sums resulted from each input bit-slice vector are aligned and buffered, quantizations happen on the BLs of buffer arrays (\circled{3}).
Digital S+A operations are still required to accumulate the digital partial sums across BLs in the buffer array (\circled{4}).} 
Such a strategy reduces the required A/D conversions to obtain the final digital dot-products.
For example, assume to use 1-bit RRAM cells to store an 8-bit weight, 
the strategy incurs $8+8-1=15$ A/D conversions in 8 input cycles, improving the energy efficiency.

Aided with the insight that performing accumulation before quantization can reduce the number of A/D conversions, we propose Strategy C in Fig.~\ref{fig:accu_compa}(c).
The idea is that if the analog partial sums from all BLs that store kernel weights can be accumulated across the input cycles, it is possible that only one A/D conversion is needed to digitize the final analog sum, minimizing the A/D energy.
Its dataflow (Strategy C in Fig.~\ref{fig:accu_compa}(d)) obeys the following steps for each input bit-slice vector $X[i]$: \circled{1} Transfer BL currents of a VMM computing array to voltages; \circled{2} Simultaneously accumulate analog partial sums (i.e., $V_0,...,V_3$ in Fig.~\ref{fig:accu_compa}(c)) across all BLs that store weights and the intermediate sum (i.e., $V_{\text{o},i-1}$) by the end of the previous input cycle $i-1$.
When the final analog sum is achieved, one-time quantization occurs (\circled{3}).
Efficient  S+A operations in the analog domain are hard to implement with conventional analog circuits~\cite{cascade}.
We thus propose a neural approximated S+A circuit (NNS+A) to enable the accumulation of analog partial sums.
As conceptually shown in Fig.~\ref{fig:accu_compa}(c), the NNS+A cyclically accumulates both the analog partial sums (i.e., $V_0,...,V_3$) of all BLs and the intermediate 
sum (i.e., $V_{\text{o},i-1}$) up until the previous input cycle $i-1$.
The temporary buffering of the intermediate sum $V_{\text{o},i-1}$ is achieved with a sample and hold (S/H) circuit as described in Section~\ref{sec: training_framework}.
In our scheme, the final analog sum is quantized by a neural approximated ADC (NNADC) reported in a prior work~\cite{pipelineadc}.
Such an NNADC can achieve high-fidelity quantization with superior energy and area efficiency. 
Additional saving is also made possible by limiting the quantization to the $P$ most significant bits (MSBs) of the final analog sum, where $P$ is the resolution of activations of a quantized DNN model.

\subsection{Unified Characterization Framework}
\label{sec:framework}

To compare the three dataflows, we build a first-order characterization framework.
It uses a few hardware parameters to derive the required A/D resolution, the number of A/D conversions, and the latency to obtain final digital dot-products for each strategy at the array level.
Without losing generality, given a pre-trained DNN model with $P_{I}$-bit inputs, $P_{W}$-bit weights, and $P_{O}$-bit outputs, the hardware parameters are defined as: 1) a VMM computing array has a size of $2^N\times2^N$ ($N$ is an integer $\leq9$); 2) each RRAM cell has a $P_{R}$-bit precision; 3) a WL is driven by a $P_{D}$-bit DAC.

\noindent\textbf{BL resolution.}
Each BL voltage in Strategy A has levels up to $(2^{P_{R}}-1)$$\times$$(2^{P_{D}}-1)$$\times $$2^N$, needing A/D resolution $P^{A}_{A}$~\cite{isaac}:
\vskip -9pt
\begin{equation}
P^A_{A}=
\left\{
             \begin{array}{lr}
              (P_{R}+P_{D}+N), ~\text{if}~P_{R}>1, P_{D}>1; &  \\
              (P_{R}+P_{D}- 1+N ), ~ \text{otherwise}.
             \end{array}
\right.
    \label{eqn:adreso_A}
\end{equation}
For Strategy B, each RRAM cell in a buffer array stores the BL information of a VMM array with the precision in Eq.~\eqref{eqn:adreso_A}.
The BL in the buffer array then needs A/D resolution of $P^A_{B}$:
\vskip -6pt
\begin{equation}
P^A_{B}= P^A_A+\log_2(\lceil P_I/P_D \rceil).
    \label{eqn:adreso_B}
\end{equation}
\textcolor{black}{Here, $\lceil \cdot \rceil$ denotes the minimal integer no smaller than ``$\cdot$''. $\lceil P_I/P_D \rceil$ is the total input cycles, i.e., the number of rows in the buffer array to store the partial sums from the VMM array.}
Strategy C extends the accumulations completely in the analog domain. 
\textcolor{black}{Though the final analog sum from the NNS+A has levels up to $(2^{P_I}-1)\times (2^{P_W}-1)\times 2^N$, we only need to extract the $P_O$ MSBs. 
}
The required A/D resolution $P^A_{C}$ is thus only determined by the output precision:
\begin{equation}
P^A_{C}= P_O.
    \label{eqn:adreso_C}
\end{equation}

\noindent\textbf{{A/D conversions.}}
For Strategy A, the total number of A/D conversions are determined by the below expression:
\begin{equation}
 N^A_{\text{Conversion}} = \lceil {P_I}/{P_D}\rceil \cdot \lceil {P_W}/{P_R} \rceil.
    \label{eqn:usage_A}
\end{equation}
\textcolor{black}{Here, $\lceil {P_W}/{P_R} \rceil$ means the total RRAM columns that store a weight. 
}
In Strategy B, only the analog partial sums of BLs in a buffer array are needed to be quantized.
The number of A/D conversions is:
\begin{equation}
 N^B_{\text{Conversion}} =\lceil 
 {P_I}/{P_D}\rceil + \lceil{P_W}/{P_R}\rceil-1.
    \label{eqn:usage_B}
\end{equation}
Since all accumulations are performed in the analog domain, only one A$/$D conversion is required by Strategy C:
\begin{equation}
 N^C_{\text{Conversion}} =1.
    \label{eqn:usage_C}
\end{equation}

\noindent\textbf{{Computation latency.}}
For all strategies, the computation cycle is determined by the input precision and the DAC resolution assuming the bit-slicing input technique.
\begin{equation}
 L^A_{\text{latency}} = L^B_{\text{latency}}= L^C_{\text{latency}}=\lceil {P_I}/{P_D}\rceil.
    \label{eqn:latency}
\end{equation}

Eq.~\eqref{eqn:adreso_A}$\sim$\eqref{eqn:latency} show the upper-bound resolutions of ADCs, the total number of A/D conversions, and the computation cycles to complete a VMM for each strategy, guiding the accuracy, energy consumption, and latency of each dataflow.
Note that the above equations are derived based on a single group of inputs and weights.
For the dot-products of a $2^N\times2^N$ VMM array, they involve multiple groups of weights.
Then, Eq.~\eqref{eqn:usage_A} to Eq.~\eqref{eqn:usage_C} should be scaled accordingly.
The characterization results at the array level for different dataflows are shown in Section~\ref{sec:cha_re}.

\subsection{Characterization Results}
\label{sec:cha_re}

We use AlexNet~\cite{alexnet} with 8-bit quantization trained on CIFAR-10~\cite{cifar} as an exemplary benchmark for the characterization.
The hardware parameters are set as follow: 1) $N=7$; 2) $P_R=1$ and $P_D\in[1,2,4]$; 3) $P_I=P_W=P_O=8$, for the array-level evaluation.
\textcolor{black}{The specifications of DACs, ADCs, digital S+As, and crossbar arrays come from the previous works~\cite{isaac, cascade} while the specifications of NNS+As and NNADCs are from Table~\ref{tb:neu_perfor} in Section~\ref{sec:performance}.}

\begin{figure}[!t]
    \centering
    \includegraphics[width=\linewidth]{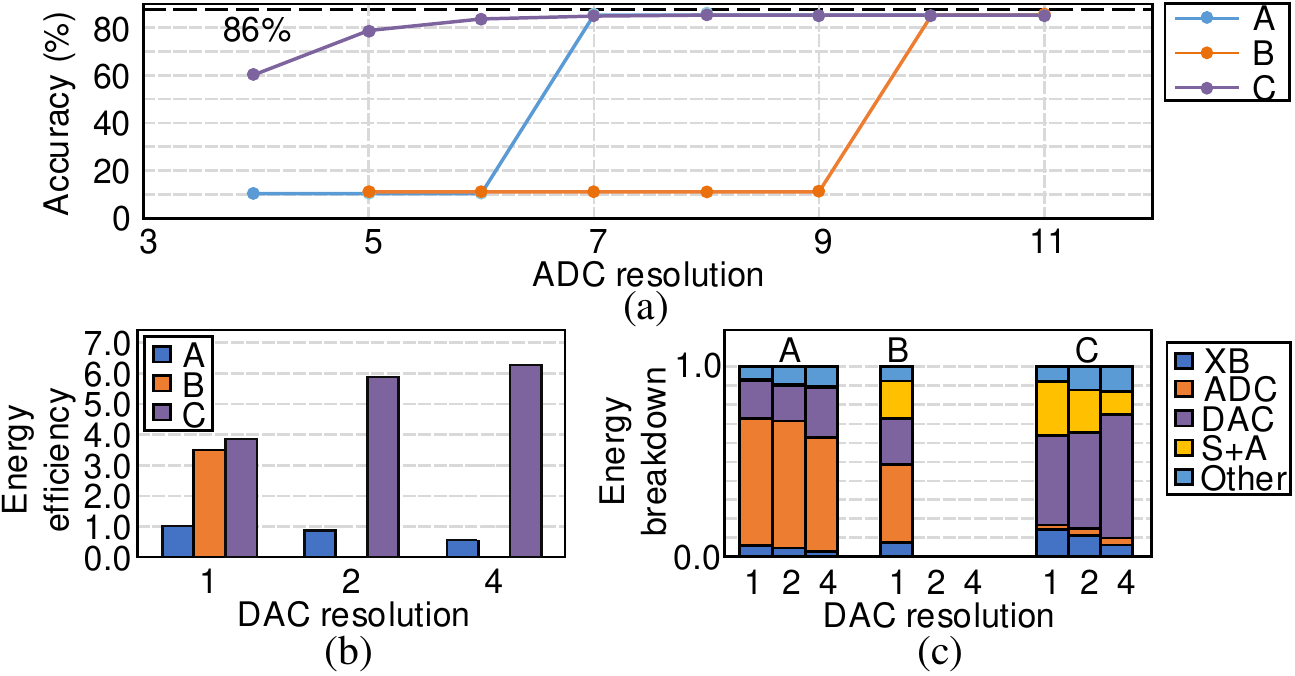}
    \caption{Dataflow comparisons. (a) Inference accuracy of AlexNet~\cite{alexnet} on CIFAR-10~\cite{cifar} as a function of A/D resolution for three strategies. (b) Normalized energy efficiency and (c) Energy breakdown of three design strategies using a 128 $\times$ 128 crossbar as an example.}
    \label{fig:characterization}
\end{figure}

\textbf{{Impact of A/D resolution on accuracy.}}
Fig.~\ref{fig:characterization}(a) shows the inference accuracy with the evolution of A/D resolutions for each strategy.
In the characterization, we set $P_R=1$, $P_D=1$.
It suggests that A/D resolutions following the theoretical bounds defined in Eq.~\eqref{eqn:adreso_A}$\sim$\eqref{eqn:adreso_C} are sufficient to deliver adequate hardware inference accuracy.
After passing the theoretical bound, the inference accuracy improvement for all three strategies is limited with the increasing of A/D resolution.
Specifically, Strategy A needs the lowest A/D resolution as it directly quantizes the BL signals and performs the digital accumulation, whereas Strategy B exhibits the most stringent requirement on A/D resolution as it buffers analog partial sums using extended RRAM buffer arrays.
Our Strategy C has a modest A/D resolution requirement as it is fully analog and the A/D resolution is only determined by the output precision $P_{O}$.
The characterizations in the remainder of this section thus use the A/D resolution dictated by the inference accuracy.

\textbf{{DAC resolution impacts.}}
In this part, we investigate how the resolution of DACs can affect the performance of the different dataflows in terms of computing dot-products.
For the investigation, we fix $P_R=1$. 
We then increase the resolution of DACs and normalize the energy consumption of all strategies with the value of Strategy A using 1-bit DACs.
As illustrated in Fig.~\ref{fig:characterization}(b), the energy efficiency of Strategy A degenerates by increasing the DACs' resolution, yet improves for Strategy C.
Although less A/D conversions are required for Strategy A based on Eq.~\eqref{eqn:usage_A}, the A/D resolution $P^A_A$ increases according to Eq.~\eqref{eqn:adreso_A} due to the higher-resolution of DACs.
The degenerated energy efficiency is caused by the exponential energy scaling law of ADC with its resolution~\cite{isaac}.
Therefore, ADC dominates the total energy consumption of Strategy A as shown in Fig.~\ref{fig:characterization}(c).
Strategy C only requires one-time A/D conversion with fixed A/D resolution $P^A_C$ determined by $P_O$.
Therefore, the A/D energy does not suffer from the increased resolution of DAC.
In addition, the computation cycles defined in Eq.~\eqref{eqn:latency} are also reduced thanks to the increased D/A resolution, improving the energy efficiency of DAC and NNS+A.
However, as DAC's power consumption scales with its resolution in a weakly exponential style~\cite{isaac, dac}, the energy efficiency of Strategy C will be dominated by DACs.
4-bit DACs are an optimal option to maximize the energy efficiency.
Only one energy breakdown of Strategy B is reported in Fig.~\ref{fig:characterization}, as the buffer array RRAM cell requires precision $>$7-bit\footnote{We follow the original scheme proposed in CASCADE~\cite{cascade} where one RRAM cell is used to buffer a high-precision analog partial sum. It may be possible (but non-trivial) to use multiple low-precision RRAM cells in a buffer array to store a high-precision analog partial sum.} when $P_D\geq2$, beyond the ability that state-of-the-art fabricated device can achieve~\cite{7bitReRAM}.
The investigation suggests that Strategy B is fundamentally limited by buffer RRAM's precision and can only adopt either low-resolution DACs or small size crossbar array for practical design considerations.

The characterizations here demonstrate that Strategy C has clear advantages in both energy efficiency and processing speed with high-resolution DACs over the other two strategies, motivating us to adopt it to improve the performance of RRAM-based PIM accelerators.

\section{NeuralPeriph Implementation}
\label{sec:nnperi}
\newcommand{\sigvtc}{\sigma_{\text{\tiny VTC}}}
\newcommand{\vcmp}{V_{\text{\tiny cmp}}}
\newcommand{\bgt}{_{\text{\tiny GT}}}
\renewcommand{\Re}{\mathbb{R}}

In this section, we present the design of the neural approximated peripheral circuits (NeuralPeriph), i.e., NNADC and NNS+A.
The detailed design of NNADCs has been proposed in our prior work~\cite{pipelineadc}. 
Hence, we give a brief introduction in Section~\ref{sec:input_range} to show its core design concept, and emphasize how it overcomes the design challenges from quantizing the analog sums with variations and variable dynamic ranges.
Here, we focus on modifying the neural approximation method~\cite{pipelineadc} to design the NNS+A.
We first introduce the hardware substrate used to instantiate a trained NNS+A model. 
We then show the offline training framework to find optimal weights for the NNS+A model to accurately approximate the ideal S+A function.

\subsection{Design of NNS+A Circuits}
\label{sec:NNS+A}

\subsubsection{RRAM-based hardware substrate}
\begin{figure}[!t]
 \centering
    \includegraphics[width=1.0\linewidth]{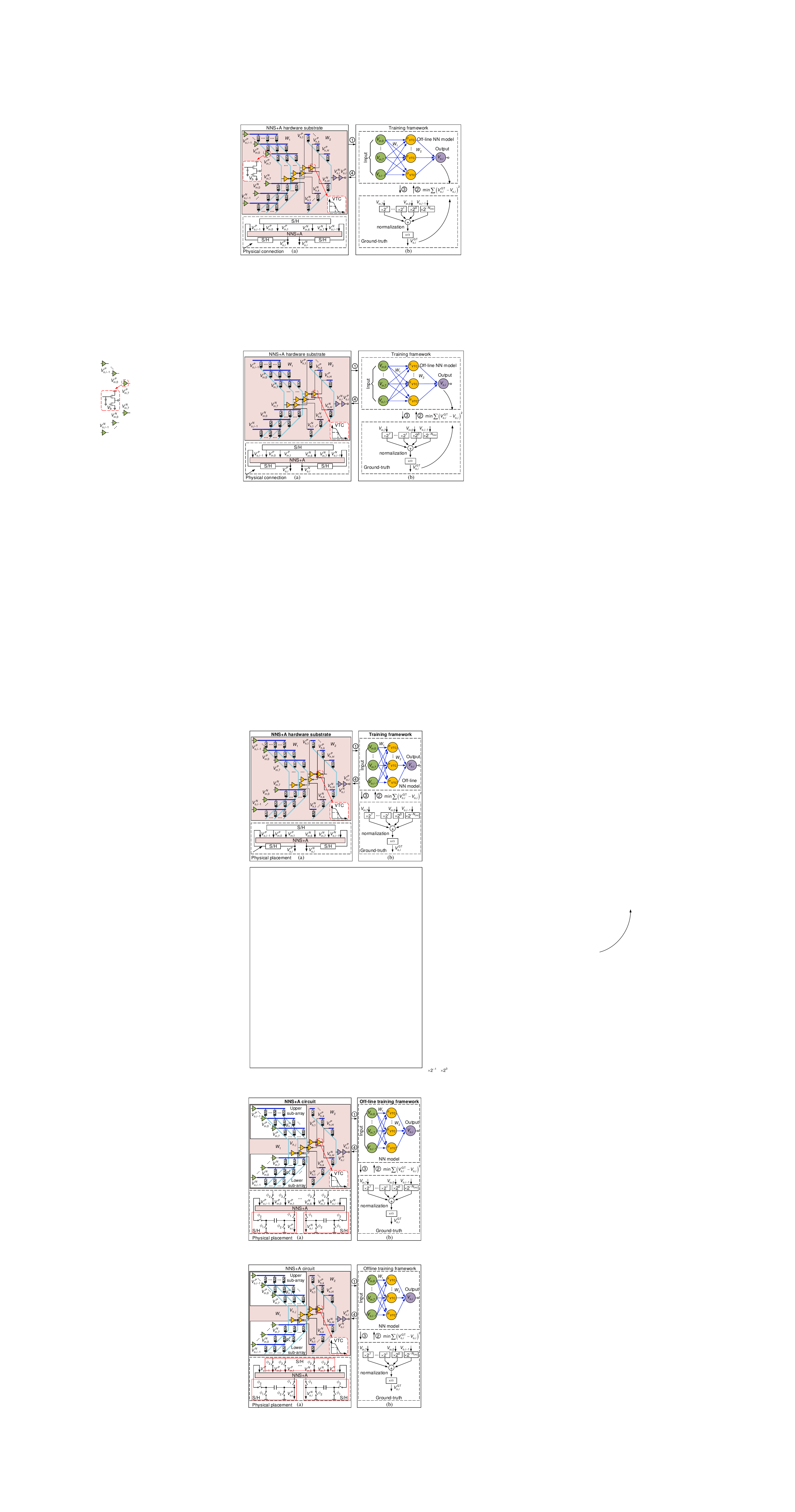}
    \caption{Design of NNS+A circuits. (a) Hardware substrate for NNS+A. (b) The offline training framework to find the optimal RRAM resistances to best approximate the S+A function.}
    \label{fig:neuralperiph}
\end{figure}

We adopt RRAM crossbar arrays and CMOS inverters as the hardware substrate~\cite{cao2019neuadc} to build the NNS+A circuit.
The RRAM crossbar arrays are used to instantiate weights and the CMOS inverters work as analog neurons\footnote{The voltage transfer characteristic (VTC) curve of a CMOS inverter preserves an S-shaped curve similar to the sigmoid and can serve as a nonlinear activation function (NAF).}.
Such a hardware substrate achieves both high energy efficiency and minimal area overhead without complex analog-style neuron circuits (e.g., operational amplifiers)~\cite{rram_approxi,cao2019neuadc}.
Fig.~\ref{fig:neuralperiph}(a) shows the NNS+A circuit with a pseudo-differential three-layer architecture.
The CMOS source followers (drivers) are used as the ``place holder'' neurons for both inputs and outputs.
Assuming that the activation precision $P_I$ and the weight precision $P_W$ of a DNN workload are 8-bit, and the RRAM cell precision $P_R$ in a VMM array is 1-bit, the NNS+A has a size of $10\times H_{\text{S+A}}\times$1.
Here, $H_{\text{S+A}}$ is the number of hidden neurons.
Among the ten pseudo-differential pairs of input ports\footnote{For the $k^{\text{th}}$ pseudo-differential pairs, two inputs satisfy the relation: $V^N_{\text{in},k}=V_{\text{DD}}-V^P_{\text{in},k}$. We assume $V^P_{\text{in},k}=V_{\text{in},k}$, then $V^N_{\text{in},k}=V_{\text{DD}}-V_{\text{in},k}$.}, eight of them are connected to the analog partial sums ($V_{\text{in},j},~j=0,1,\cdots,7$) from the BLs that store 8-bit weights.
The $9^{\text{th}}$ one is connected to the intermediate analog sum ($V_{\text{o},i-1},~i=0,1,\cdots,8/N_{\text{DAC}}-1$) utill the previous cycle $i-1$. 
The remaining one is kept for the trained biases (connected to $V_{\text{DD}}$ or GND) which are not explicitly shown in the Fig.~\ref{fig:neuralperiph}(a).

Taking the first layer of the NNS+A as an example, the crossbar array achieves VMM as
$V_{\text{h},j} =\sum\nolimits_{k=1}^{9} W_{k,j} \cdot V_{\text{in},k} + V_{\text{off},j},j\in{1,2,...,H_{\text{S+A}}},$
with $k$ and $j$ as the indices of input/output ports.
The bipolar weight $W_{k,j}$ is the differential of two cell conductances in the upper ($U$) sub-array and the lower ($L$) sub-array, which is defined as
\begin{equation}
\label{eq:cross_w}
W_{k,j}=\epsilon\cdot \Big(g^U_{k,j} - g^L_{k,j}\Big), ~~\epsilon=1/\sum\nolimits_{k=1}^{H_\text{S+A}} \Big(g^{U}_{k,j}+g^{L}_{k,j}\Big).
\end{equation}
Therefore, once the weights of the NNS+A model in Fig.~\ref{fig:neuralperiph}(b) are trained, they can be instantiated as the conductances of RRAM devices in the hardware substrate using Eq.~\eqref{eq:cross_w}.

\subsubsection{Offline training framework}
\label{sec: training_framework}
The proposed training framework for NNS+A circuit is shown in Fig.~\ref{fig:neuralperiph}(b).
It can accurately capture the circuit-level behaviors of the hardware substrate and learn the associated hardware design parameters (i.e., RRAM conductances) by approximating the ideal input/output relationship of S+A.
There are four essential steps in the training framework.

In Step \circled{1}, the NNS+A circuit in Fig.~\ref{fig:neuralperiph}(a) is modeled as a three-layer NN in Fig.~\ref{fig:neuralperiph}(b) with a single hidden layer:
\begin{equation}
\label{eq:target}
  \tilde{h} = L_1(\vec{V}_{\text{in}}; \theta_1);~~~V_{\text{h}} = \sigvtc(\tilde{h});~~~V_{\text{o}} = L_2(V_{\text{h}};\theta_2).
\end{equation}
Here, $\vec{V}_{\text{in}}$ is an analog input vector and $V_{\text{o}}$ is an analog output.
$\tilde{h}$ denote voltages at the output of the first crossbar layer.
They are modeled as a linear function $L_1$ of $\vec{V}_{\text{in}}$ with learnable parameters $\theta_{1}=\{W_1,V_1\}$, corresponding to the weights and bias associated with the first layer.
Each of these voltages passes through an inverter, whose input-output relationship is modeled by the nonlinear function $\sigvtc(\cdot)$, to yield the vector $V_{\text{h}}$. 
The linear function $L_{2}$ models the second crossbar layer to produce the output $V_{\text{o}}$ with learnable parameters $\theta_{2}=\{W_2,V_2\}$, corresponding to the weights and bias associated with the second layer.
To thoroughly model the hardware behavior using Eq.~\eqref{eq:target}, three practical hardware constraints are considered: PVT variations of CMOS neurons, limited precision and variation of RRAM devices, and constrained weight value:
\begin{equation}
\label{eq:w_cons}
\sum(\text{abs}(W_1),0)<1;~~  \sum(\text{abs}(W_2),0)<1,
\end{equation}
caused by the passive crossbar array shown in Eq.~\eqref{eq:cross_w}.
Step \circled{4} gives the details to incorporate them into training.

In Step \circled{2}, a learning objective is established to find optimal $\{\theta_{1},\theta_{2}\}$ (associated with RRAM  conductances in crossbar arrays) such that for all values $\vec{V}_{\text{in}}$ in the input range, the circuit yields corresponding output $V_{\text{o}}$ that are equal to or close to the desired ``ground-truth'' $V_{\text{o,GT}}$.
Therefore, the learning objective is formulated as $\{\theta_{1},\theta_{2}\}= \arg \min \sum\nolimits_{t}  C_t(V_{\text{o}},V_{\text{o,GT}})$.
Here, the cost function $C$ is defined as the mean-square error (MSE), measuring the discrepancy between the predicted $V_{\text{o}}$ and true $V_{\text{o,GT}}$, e.g., $ C(V_{\text{o}},V_{\text{o,GT}}) = \sum\nolimits_{k} ({V_{\text{o,GT}}(k)}-V_{\text{o}}(k))^2$.

In Step \circled{3}, ``ground-truth'' datasets are generated for training.
The NNS+A circuit accumulates the analog partial sums of input bit-slice vectors and a group of weights.
Its input/output mapping relationship, i.e., the ``ground-truth'', is modeled as $V_{\text{o,GT}}=\big({2^{-N_{\text{DAC}}}\cdot V_{\text{o},i-1} + \sum\nolimits^{7}_{j=0}2^{j}\cdot V_{\text{in},j}}\big)/{\alpha}$.
Here, $V_{\text{in},0},\cdots,V_{\text{in},7}$ are the ideal analog partial sums in each input cycle from the BLs that store 8-bit weights, and $V_{\text{o},i-1}$ is the ideally intermediate sum until the previous cycle $i-1$.
We assume the LSBs of inputs are first streamed into the VMM computing array.
Hence, the $V_{\text{o},i-1}$ is multiplied by a coefficient $2^{-N_{\text{DAC}}}$. 
We normalize the ideal output by $\alpha=2^{-N_{\text{DAC}}}+\sum\nolimits^{7}_{j=0}2^{j}$ to keep it in the same range as the input signals.
Prior work~\cite{isaac,cascade} shows that the analog partial sum generated by the MSBs of inputs and weights contributes most to the MSBs of the final digitized sum.
This LSB-first streaming method can thus reduce the computation accuracy loss due to the incomplete charge transfer from the repeated accumulations, as $V_{\text{o},i},~i=8/N_{\text{DAC}}-1$ (i.e., partial sum generated by MSBs of inputs and weights) needs one-time accumulation.
Additionally, it also helps to minimize the computation errors in each input cycle as these computation errors are attenuated by the normalization factor $\alpha$ ($\alpha>1$). All these errors are included in our dataflow accuracy analysis in Section~\ref{sec:accu_ana}.

In Step \circled{4}, we leverage hardware-aware training techniques to find the feasible and robust weights for the trained NN model.
At the the beginning, we initialize the parameters $\{\theta_{1}, \theta_{2}\}$ randomly, and update them iteratively based on the gradients computed on the mini-batches of $\{\vec{V}_{\text{in}},V_{\text{o,GT}})\}$ pairs, which are randomly sampled from the input range.
To incorporate the hardware constraints in Step \circled{1} into training, we let each neuron $j$ in Eq.~\eqref{eq:target} randomly select a VTC from $A_{\text{VTC}}$ which is a group of VTCs with different PVT variations during training: $\sigvtc^j = A_{\text{VTC}}[f_\text{randint}(N_{\text{VTC}})]$.
Here, $f_\text{randint}(N_{\text{VTC}})$ is a function to generate a random integer  $<N_{\text{VTC}}$.
We then set $W_i, i=1,2$ as $A_{\text{R}}$-bit to meet the requirement of RRAM precision, and perturb $W_i$ using $W_i \leftarrow W_i\cdot e^{\theta},~\theta \sim \mathcal{N}(0,\sigma)$ to reflect the stochastic variation of RRAM resistance~\cite{cao2019neuadc}, and periodically clip all values of $W_1$ ($W_2$) between $[-1/10,1/10]$ ($[-1/H_{\text{S+A}},1/H_{\text{S+A}}]$) to satisfy Eq.~\eqref{eq:w_cons} in the training.
We also add Gaussian noise into the ground-truth inputs to mimic the thermal noise of S/H circuits. 
After training, we adopt the same method in the previous work~\cite{cao2019neuadc} to instantiate the RRAM conductance for the NNS+A circuit from the trained weights.
With the Step \circled{1}$\sim$\circled{4}, a robust and accurate NNS+A circuit can be implemented, as the SPICE simulation results shown in Table~\ref{tb:neu_perfor}. 
To ensure the proposed NNS+A circuit accumulate the intermediate analog sums cyclically, its outputs are connected to the $9^{\text{th}}$ inputs pair via S/H circuits~\cite{sh} shown at the bottom of Fig.~\ref{fig:neuralperiph}(a).
The differential clocks control $\phi_1$ and $\phi_2$ alternatively to sample and hold (or transfer) the charge, thereby buffering the intermediate sum $V_{\text{o},i-1}$.

\begin{figure}[!t]
    \centering
        \includegraphics[width=1\linewidth]{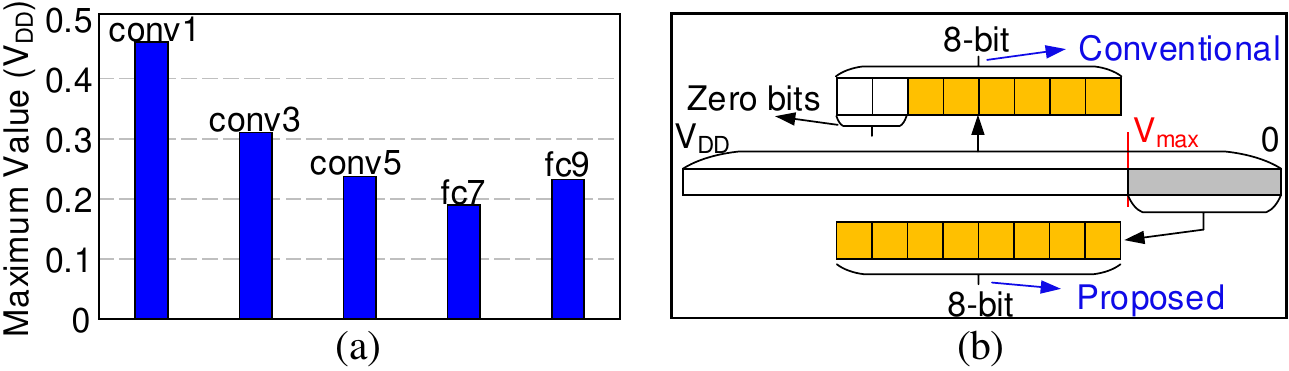}
    \caption{(a), Distribution of maximum voltages of NNS+A for different layers in Alexnet~\cite{alexnet}. 
    (b), Quantizing NNS+A’s output using its dynamic range $[0,V_{\text{max}}]$.}
    \label{fig:weight_distribution}
\end{figure}

\subsection{Input Range-Aware Training for NNADC Circuits}
\label{sec:input_range}
Our NNADC is implemented based on a pipelined structure with the same hardware substrate and similar design procedures as the NNS+A.
The main difference is that its ``ground-truth'' is the quantization function of an ADC.
The design details are discussed in a prior work~\cite{pipelineadc}.
Here, we introduce several key challenges required to be overcome when designing an NNADC.
As discussed in Section~\ref{sec:charac},
NNADCs with 8-bit resolution are sufficient for the quantization of 8-bit DNN models.
An implicit assumption there is that the output of an NNS+A could span the fixed full range of $[0,V_{\text{DD}}]$.
However, since the weights and activations of DNN layers usually exhibit normal distributions, we observe that BLs' outputs of a VMM computing array are usually smaller than $V_{\text{DD}}$.
It suggests that the range of the final analog sums from an NNS+A is less than $V_{\text{DD}}$ and this range can be different across the DNN layers. 
Fig.~\ref{fig:weight_distribution}(a) illustrates a distribution of the maximum output voltage of ideal NNS+A circuits for different layers in AlexNet~\cite{alexnet}.
The conventional quantization scheme based on a full-scale range of $[0,V_{\text{DD}}]$ is ineffective, as several MSBs of the digital codes may be zero (Fig.~\ref{fig:weight_distribution}(b)), degenerating the inference accuracy.
Moreover, despite the proposed mitigation techniques, the final analog sums from a practical NNS+A circuit still suffer from its inherent noise,
which further deteriorates the hardware inference accuracy.

To resolve these challenges, we propose an input range-aware technique to design NNADCs (Fig.~\ref{fig:weight_distribution}(b)) by defining the labeled digits using the dynamic range $[0, V_{\text{max}}]$ of ideal NNS+A circuits' final output $V_{\text{ideal}}$:
\begin{equation}
\label{eqn:adc}
\sum_{i=0}^{7} 2^{i} \cdot d_{i} = \text{round}\left( {V_{\text{ideal}}}/{V_{\max}} \times (2^{8}-1)  \right).
\end{equation}
Here, ``$\text{round}(\cdot)$'' is to get the closest integer of ``$\cdot$''; 
$d_i$ is the $i^{\text{th}}$ digit of an 8-bit digital code associated with the $V_{\text{ideal}}$,
whereas the inputs used to train NNADCs come from the  final sums $V_{\text{o}}$ of practical NNS+A circuits.
Each value of $V_{\text{o}}$ is a noisy version corresponding to its ideal value $V_{\text{ideal}}$. 
We therefore can compensate for the errors caused by the non-idealities of NNS+A circuits by using the noisy inputs ($V_{\text{o}}$) with the correct labels (Eq.~\eqref{eqn:adc}) to train the NNADCs.
We find that training three NNADCs by letting $V_{\text{max}}$ in Eq.~\eqref{eqn:adc} be $0.5V_{\text{DD}}$, $0.25V_{\text{DD}}$, $0.125V_{\text{DD}}$ provides sufficient coverage for various DNN models.
For each DNN layer, we then program the hardware substrate of NNADCs to be one of the three pre-trained NNADC models to accommodate the dynamic output ranges and noisy outputs of NNS+A circuits.
In this way, most significant information of NNS+A's output is preserved by the 8-bit quantization.

\begin{table}[!tb]
    \footnotesize
\centering
\caption{Performances of trained NeuralPeriph circuits.}
\begin{threeparttable}
\begin{tabular}{c|c|c|c|c}
\toprule
NeuralPeriph                                                                          & \multicolumn{2}{c|}{NNS+A}            & \multicolumn{2}{c}{8-bit NNADC}     \\ \midrule
Supply                                                                            & \multicolumn{4}{c}{1.2 $V$}                                               \\ \hline
Technology                                                                      & \multicolumn{4}{c}{130 nm}                                               \\ \hline
RRAM                                                                 & \multicolumn{4}{c}{precision: $A_{\text{R}}=3$-bit;~  variation: $\sigma=0.025$}                                               \\ \hline

Input range                                                                        & \multicolumn{4}{c}{$[0,0.5]$ $V$}                                             \\ \hline
\multirow{2}{*}{Speed}                                                                & \multicolumn{2}{c|}{MHz}              & \multicolumn{2}{c}{GS$/$s}      \\ \cline{2-5} 
                                                                                                & 20          & 40        & 0.5           & 1             \\ \hline
Area ($\text{mm}^2$)                                                                                & $1.5\cdot10^{-3}$     & $3\cdot 10^{-3}$    & 0.0069         & 0.015         \\ \hline
Power (m$W$)                                                                                   & 0.68        & 1.39      & 6.3            & 13.1          \\ \hline
ENOB (bits)                                                                           & \multicolumn{2}{c|}{N$/$A}              & 7.88           & 7.85          \\ \hline
\multirow{4}{*}{\begin{tabular}[c]{@{}c@{}} \\Approximation \\ error\\ \end{tabular}} & \multicolumn{2}{c|}{Max error (m$V$)} & \multicolumn{2}{c}{DNL (LSB)\tnote a} \\ \cline{2-5} 
                                                                                                & 4           & 5         & $-0.25/0.55$      & $-0.23/0.5$     \\ \cline{2-5} 
                                                                                      & \multicolumn{2}{c|}{Min error (m$V$)} & \multicolumn{2}{c}{INL (LSB)\tnote a} \\ \cline{2-5} 
                                                                                              & $-$3          & $-$4        & $-0.56/0.62$      & $-0.45/0.6$     \\ \bottomrule
\end{tabular}
\begin{tablenotes}\scriptsize
\item[a] A normal range for the DNL and INL of conventional ADCs is $(-1,1)$LSB (least significant bit).
The smaller this range is, the better the NNADC is.
\end{tablenotes}
\end{threeparttable}
\label{tb:neu_perfor}
\end{table}

\subsection{Evaluation of NeuralPeriph Circuits}
\label{sec:performance}
We use the configurations introduced in Section~\ref{sec:ea_model} to train a group of NeuralPeriph circuits.
Our offline training framework yields high accuracy for the trained NeuralPeriph circuits.
For example, an NNS+A model can achieve $<10^{-5}$ MSE.
The NeuralPeriph circuits are thus considered without explicit approximation errors.
We then focus on evaluating the effect of their inherent hardware non-idealities on the computational accuracy.
The circuit-level simulations are presented in Table~\ref{tb:neu_perfor}.
We use two important metrics of conventional ADCs, i.e., differential non-linearity (DNL) and integral non-linearity (INL), to evaluate the computation accuracy of NNADCs, and use the absolute error between the ideal output $V_{\text{o},\text{GT}}$ and the simulated circuit-level output $V_\text{o}$ to evaluate the computation accuracy of NNS+A.
With the hardware-aware training technique proposed in Section~\ref{sec: training_framework}, the circuits demonstrate high computation accuracy.
Their performance, e.g., speed and power, conservatively obtained from SPICE simulation, is also listed in Table~\ref{tb:neu_perfor}, exhibiting high energy efficiency.

\section{Neural-PIM Accelerator}
\label{sec:architecture}

This section discusses the design of Neural-PIM accelerator.
We starts by describing its overall architecture.
We then elaborate the system implementation.
Finally, we perform an analysis on its inference accuracy.

\subsection{Architecture Overview}
\label{sec:overll_archi}

\begin{figure}[!t]
 \centering
    \includegraphics[width=1.0\linewidth]{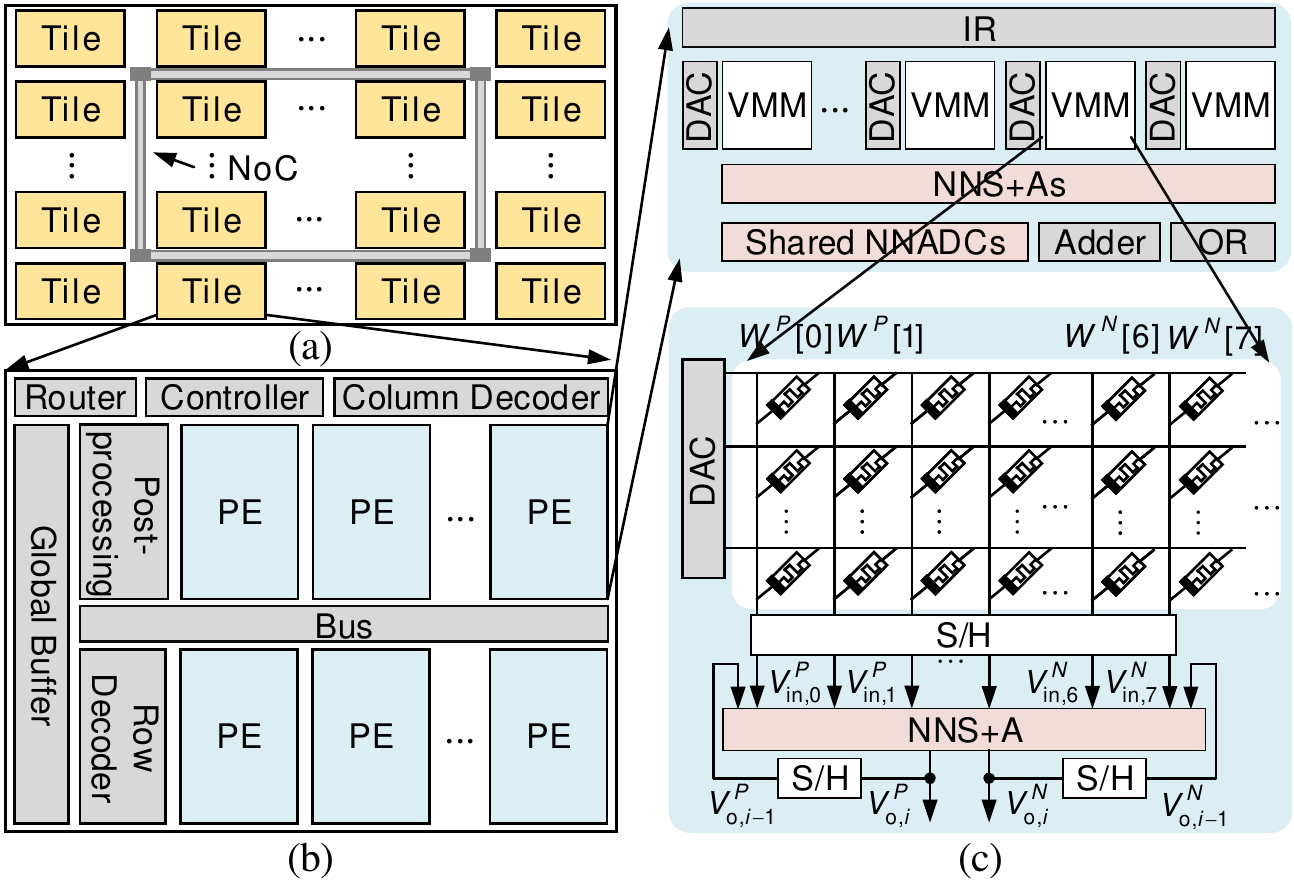}
    \caption{Overview of the Neural-PIM architecture. (a), Floorplan of the accelerator.
    (b), The organization of a processing tile. (c), Deployment of NNS+As in PEs and their physical connections with VMM arrays.}
    \label{fig:overallarch}
\end{figure}

The proposed Neural-PIM accelerator aims to improve the energy efficiency and performance for DNN inference by adopting the proposed analog accumulation scheme in Fig.~\ref{fig:accu_compa}(c).
Fig.~\ref{fig:overallarch}(a) illustrates its architecture overview. 
It consists of several tiles which are connected with network-on-chip (NoC).
In each tile (Fig.~\ref{fig:overallarch}(b)), there are digital components, such as buffers, input/output registers (IRs/ORs), decoders, controllers, and post-processing units, and analog/mixed-signal components, such as processing elements (PEs).
The digital components are necessary to provide data storage, address decoding, processing controller, and light-weight computation functionalities (e.g., activation function, pooling, element-wise multiplication, and aggregation) for the accelerator; 
while the PEs, each consisting of a number of VMM computing arrays and NNS+A circuits, perform the most computation-intensive multiplication-and-accumulation operations.
The NNADCs quantize the accumulated analog sums into digital formats for post-processing.
Since the number of A/D conversions are minimized, several NNADCs are shared by all PEs in contrast with previous work~\cite{isaac,enabling_sci} that requires a dedicated ADC for each RRAM crossbar array.

The accelerator supports DNN models with 8-bit quantization.
After offline training, the weights of each DNN are programmed onto the memristors. 
NeuralPeriph circuits are similarly instantiated on the RRAM-based hardware substrate once their models are trained\footnote{In contrast to the buffer arrays in CASCADE~\cite{cascade} where frequent write operations to RRAM cells are necessary, our NeuralPeriph circuits can be instantiated by writing to their RRAM cells once and then function as inference modules with read-only operations.}.
Control vectors then are loaded into each tile to drive the finite state machines that coordinate inputs and outputs correctly every cycle.
The accelerator executes a DNN model in a pipelined manner (Section~\ref{sec:pipeline}).
Each RRAM crossbar array has a size of $128\times 128$ where each RRAM cell has 1-bit precision.
We leverage the bit-slice technique for input streaming, that is to use $N_{\text{DAC}}$-bit DAC to sequentially send $N_{\text{DAC}}$ bits of 8-bit inputs into the WL per input cycle.
We leave $N_{\text{DAC}}$ as a hyper-parameter for design space exploration in Section~\ref{sec:exh}.

\subsection{System Implementation}
\label{sec:system}

\subsubsection{Weight mapping and input streaming}

DNN weights $W$ are usually signed value while RRAM conductances are positive.
Previous work~\cite{isaac, enabling_sci,pipelayer,prime, sparse_ReRAM,cascade} decomposes signed weights $W$ into the subtraction of two positive weights, i.e., $W=W^P-W^N$, and uses two sets of crossbar arrays, i.e., positive crossbar array (for $W^P$) and negative crossbar array (for $W^N$) to store them.
After VMMs are finished in crossbar arrays, the analog partial sums of two corresponding BLs in the two arrays are subtracted by an analog subtraction unit, i.e., $W\cdot X=(W^P-W^N)\cdot X$.
In contrast with previous RRAM-based PIM accelerators~\cite{isaac, enabling_sci,pipelayer,prime,sparse_ReRAM,cascade}, we store $W^P$ and $W^N$ into the adjacent columns of the same crossbar array using 1-bit RRAM cells, as illustrated in Fig.~\ref{fig:overallarch}(c).
This weight mapping method is compatible with the proposed NNS+A circuit, which will be discussed in Section~\ref{sec:dep_neu}.
In particular, for an 8-bit DNN model, a $128\times$128 array stores 8 weights per row and 1024 weights in total.
Every group of 16 columns corresponds to a $128\times1$ 8-bit weight vector.
A large kernel can be stored into multiple crossbar arrays and even PEs and Tiles.
In this case, some of the tiles will be designed as aggregators, i.e., they aggregate ORs of different tiles and apply them in activation units for the next layer.

\subsubsection{Deployment of neural-approximated peripherals}
\label{sec:dep_neu}

The NeuralPeriph circuits are designed using crossbar arrays and CMOS inverters.
They can be naturally be deployed in RRAM-based PIM accelerators with minimal area overheads.
Fig.~\ref{fig:overallarch}(c) shows the detailed deployment of NeuralPeriph circuits in the proposed accelerator.
The BLs that store $W^P$ and $W^N$ form pseudo-differential pairs, whose outputs are connected to the associated input ports of an NNS+A circuit.
The NNS+As are placed below VMMs.
By storing $W^P$ and $W^N$ into the same array, the length of wires connected to the NNS+A circuit is reduced, minimizing the latency and parasitics.
As mentioned in Section~\ref{sec:NNS+A}, the outputs of NNS+As are connected to the $9^{\text{th}}$ inputs pair via S/H circuits to achieve the analog accumulation of partial sums cycle by cycle.
Note that each group of $128\times1$ 8-bit weight vector uses an NNS+A, and several NNS+As can be shared by a $128\times128$ crossbar array depending on their operating speed.
The NNADCs are placed at the center between PEs in order to minimize parasitics.
Their inputs are connected to the outputs of NNS+As via multiplexers.
After the end of analog accumulation, the NNADCs perform quantizations.

\subsubsection{Local buffers and digital processing units}
\label{sec:buf_digital}
In Neural-PIM, we make use of memory buffer and IRs/ORs in the memory hierarchy, which is similar to the previous RRAM-based PIM accelerators~\cite{isaac,cascade}.
The memory buffer stores a large amount of inputs/activations that are sent into the tile or generated by the tile.
IRs/ORs serve as its cache.
In particularly, IR stores part of input/activations sent to DACs while OR stores the quantized dot-products from NNADCs.
We adopt the eDRAM as the memory buffer~\cite{isaac,cascade} and employ SRAM as the IRs/ORs whose capacities are designed to accommodate the input/output data rate.
The hardware overhead of local buffers can be reduced
by buffer sharing of all PEs. 
IRs/ORs are shared by all RRAM crossbar arrays in one PE while the memory buffer is shared by all PEs.
The dot-products stored in ORs are sent to post-processing units for further processing.
In this stage, the element-wise operations of RNN (the last two rows in Eq.~\eqref{eqn:lstm}), digital activation ($\tanh$, sigmoid for RNN, and ReLU for CNN), and pooling are performed. 
The finally processed results will be stored into memory buffer for the processing of the next layer.

\begin{figure}[!t]
    \centering
        \includegraphics[width=1\linewidth]{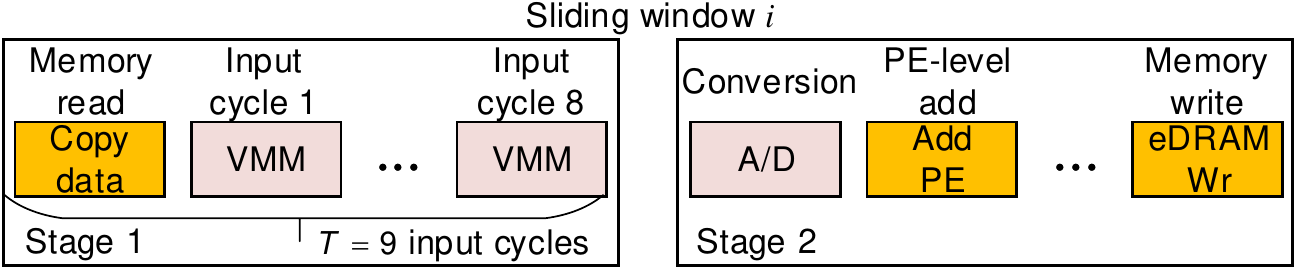}
    \caption{A coarse-grained pipeline for CNNs in a Tile.}
    \label{fig:pipeine}
\end{figure}

\subsubsection{Coarse-grained pipeline between tiles and NoC}
\label{sec:pipeline}
CNNs are composed of multiple cascaded layers, such as CONV, POOL, and FC. 
The sliding window operation scheme allows a pipelined manner to run CNN workloads on the hardware.
A coarse-grained Tile pipeline for CNN inference is illustrated in
Fig.~\ref{fig:pipeine}, which includes two stages.
In the fist stage, PEs will be busy with analog VMM operations after copying input data of a sliding window  from eDRAM into the IRs.
In the second stage, main digital operations are required to be performed after quantizations, such as PE/Tile level accumulation, digital activation, and storing activations into eDRAM to provide the inputs for the next CONV.
POOL operation is also contained in this stage.
These digital operations are executed sequentially.
Meanwhile, the first stage begins to process another sliding window.
Note that eDRAM read and write are separated into the two stages to avoid structure hazards.
Each pipeline cycle takes up 9 input cycles and each input cycle is $100$ ns as proposed by the previous work~\cite{isaac}.

Usually, the sliding window in a CONV shifts towards right by strides $S_x$ or down by strides $S_y$.
For example, if $S_x=2$ and $S_y=1$, the previous CONV layer $i-1$ needs to produce two values such that the current CONV layer $i$ can perform its next step.
This stride with value larger than one can cause an unbalanced pipeline in the proposed Neural-PIM accelerator, that is the when the PEs of layer $i-1$ are busy in every cycle, the PEs of layer $i$ has to stop for one cycle.
To make the pipeline balanced, we adopt the weights replication strategy proposed in previous work~\cite{isaac}.
For the previously given case, we double the resources allocated to layer $i- 1$, that is the weights for layer $i-1$ are replicated in different crossbar arrays in order to process two different input vectors in parallel such that two output values are produced in one cycle.
Relying on the values of $S_x$ and $S_y$ of each layer, the weights in early layers may be replicated accordingly.
However, it should be noted that the aggregated storage requirement of replicating weights should be in the range of the available storage on the chip.

RNNs consist of only fully connected layers followed by element-wise operations (EM).
Their executions can also follow a two-stage pipeline similar to CNNs.
In the first stage, analog VMMs are performed while
in the second stage, digital operations such as activation and EM operations are conducted.
Unlike CNN workloads, weight replications are not required in RNN workloads.
However, depending on the number of crossbars required by different RNN models, tiles can be grouped into multiple channels, each of which implements the whole LSTM processing function for one input vector sequence independently.

NoC is a key part of accelerator which contributes to the energy consumption and area overhead.
We adopt a similar NoC implementation proposed in a prior work~\cite{era-lstm}.
We use a concentrated mesh (c-mesh) as the NoC topology to reduce the hardware overhead of communication, where routers are shared among adjacent Tiles.

\subsection{System Accuracy Analysis}
\label{sec:accu_ana}

\subsubsection{Noise characterization of the analog dataflow}
\label{sec: noise_char}
Our proposed Neural-PIM accelerator employs an analog dataflow for accumulation, and hence is subject to various hardware non-idealities, such as RRAM read noise\footnote{Including RRAM non-idealities in both NeuralPeriph circuits and VMM computing arrays.} and PVT variations, thermal noise and bias caused by incomplete charge transfer of the S/H circuits.
Together, these non-idealities act to influence the accuracy of the Neural-PIM accelerator.
It is thus critical to examine the noise tolerance of the analog dataflow.
To tackle this issue, we develop an error model for the analog dataflow.
A final digital-dot product $D_{\text{hw}}$ (e.g., output from an NNADC) produced by hardware with the analog dataflow can deviate from its ideal value $D_{\text{sw}}$ computed by software.
The difference is modeled as $D_{\text{hw}} = D_{\text{sw}} + \mathcal{N}(0,\epsilon)$.
Here, $\mathcal{N}(0,\epsilon)$ is a Gaussian model for the lumped noise\footnote{The lumped noise includes all hardware non-idealities of the analog dataflow, e.g, variations of reading RRAM, PVT variations of NeuralPeriph circuits, and thermal noises and incomplete charge transfer of S/H circuits.}  in the whole analog dataflow (starting from the D/A conversion and ending after the A/D conversion).

To characterize the lumped noise of the analog dataflow, we perform SPICE simulations in an end-to-end manner.
We choose a kernel with random weights and map them into the hardware.
By sourcing a group of random inputs into the hardware through DACs, we obtain the practical digital outputs $D_{\text{hw}}$ from NNADCs and then compare them with their ideal outputs $D_{\text{sw}}$ to get the variation.
Performing hundreds of such Monte Carlo (MC) simulations, the statistics of the errors can be obtained.
The variation $\epsilon$ of the Gaussian noise is then expressed as $(\epsilon)^2=P_{\text{noise}}=\text{mean}(D_{\text{hw}}-D_{\text{sw}})^2$.
With the variation, the signal-to-noise and distortion ratio (SINAD) of the analog dataflow is characterized as $\text{SINAD}_{\text{hw}} = 10\log10\Big(\big({P_{\text{sig}}(D_{\text{sw}})+P_{\text{noise}}}\big)/{P_{\text{noise}}}\Big)$.
We use this SINAD as the metric to evaluate the accuracy of the analog dataflow.
Intuitively, the higher the SINAD is, the more accurate the computation is.

\begin{figure}[!t]
    \centering
    \includegraphics[width=1\linewidth]{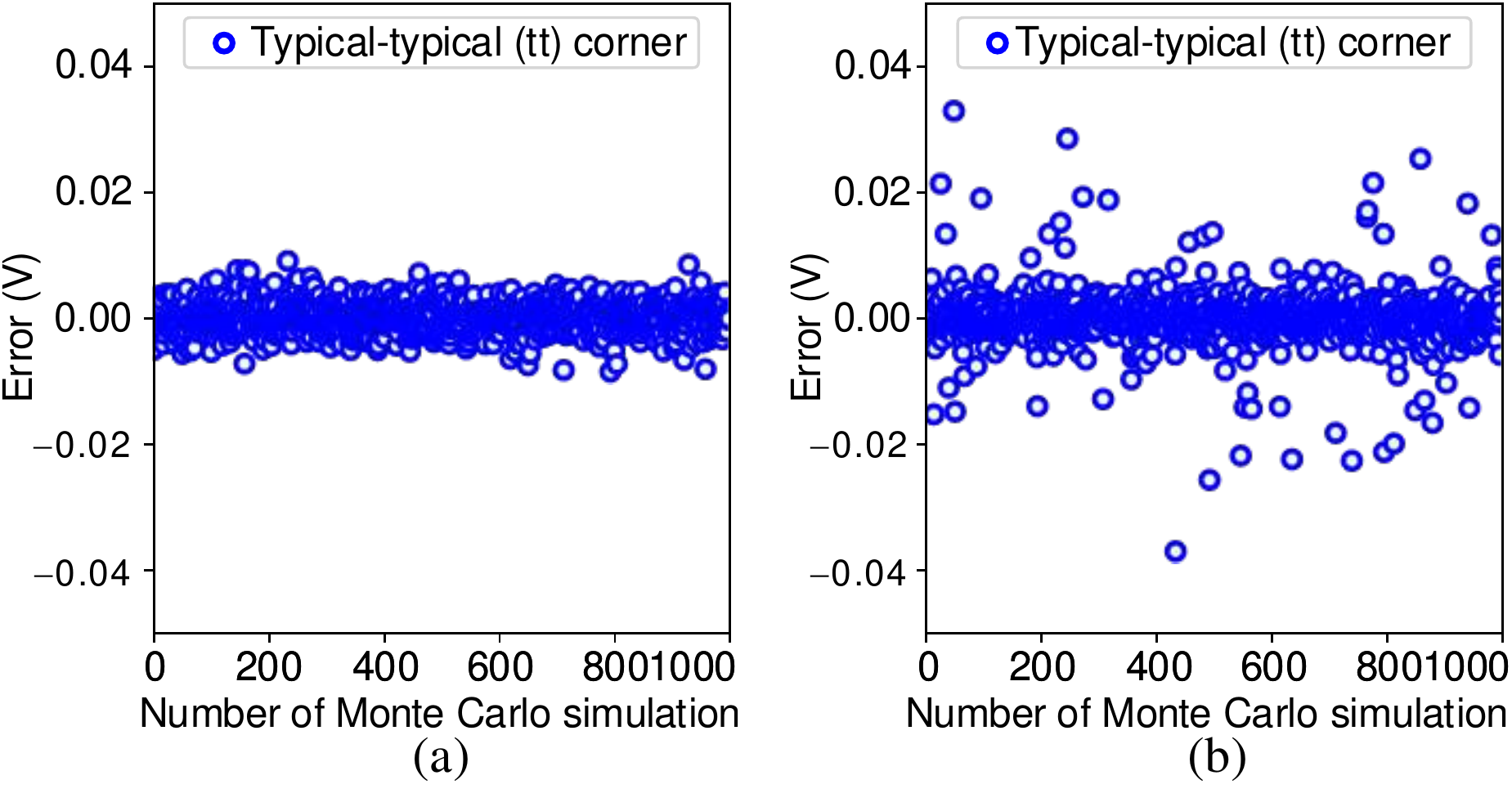}
    \caption{The difference between circuit-level simulated output $D_{\text{hw}}$ of a DNN layer with NeuralPeriph circuits and the ideal output $D_{\text{sw}}$. (a)/(b), With/without circuit-level optimization techniques. 
    We run 1000 Monte Carlo simulations at typical-typical (tt) corner.}
    \label{fig:noise_sim}
\end{figure}

Fig.~\ref{fig:noise_sim}(a) demonstrates the differences between $D_{\text{hw}}$ and $D_{\text{sw}}$ using 1000 MC simulations.
The computation errors are generally in the range of $[-0.01,0.01]V$, equivalent to a high SINAD value of 50 dB for the end-to-end analog accumulation.
It is made possible by a number of circuit-level techniques discussed in Section~\ref{sec:nnperi}.
First, our neural approximation design methodology incorporates RRAM device variations and CMOS PVT variations into the training process to enhance the robustness of the NeuralPeriph circuits against these non-idealities.
Second, the streaming order of high-precision inputs is deliberately chosen to be LSB-first to alleviate the impact of incomplete charge transfer on the precision of the analog partial sum.
Finally, to further compensate for the error due to repeated accumulations, we use the actual outputs from the non-ideal NNS+As as ground-truth inputs and the corresponding digitized values of their ideal outputs as the labels to train the NNADCs.
As a comparison, we also perform SPICE simulations to obtain the computation errors of the analog dataflow without these optimization techniques.
Fig.~\ref{fig:noise_sim}(b) shows the computation errors using the same number of MC simulations in Fig.~\ref{fig:noise_sim}(a)--the errors have grown to $[-0.04,0.04]V$, decreasing the SINAD to 35 dB.

\subsubsection{Accuracy characterization of the system}
\label{sec:accu_cha}

To examine whether the SINAD derived from our analog dataflow can support adequate system-level inference accuracy, we analyze the accuracy of a DNN model on Neural-PIM by sweeping the level of SINAD.
In this way, we can obtain $\text{SINAD}_{\min}$, the minimum  SINAD required to achieve the software-equivalent inference accuracy\footnote{If $\text{SINAD}\geq \text{SINAD}_{\min}$, we consider that the analog dataflow has sufficient computation accuracy to guarantee the system-level inference.}.
According to the prior work~\cite{cascade}, the effect of the hardware-level noise can be modeled as additive Gaussian noise to the ideal activations of DNN layers.
We thus adopt a similar approach to perform the accuracy characterization.
Experimentally, the noise injection is formulated in the following manner. 
Let $x_i$ be the output activations of a CONV layer $i$ or a FC layer $i$ (taking CNN as an example), the noise level corresponding to an $\text{SINAD}$ that would be injected into the activations is
\begin{equation}
    \sigma_i = {\max |x_i|}/{10^{\frac{\text{SINAD}}{20}}}.
       \label{eqn:s_snr}
\end{equation}
The practical activations used at the software-level inference are then given by $x^{'}_i = x_i + \mathcal{N}_{\text{inj}}(0, \sigma_i)$, where $\mathcal{N}_{\text{inj}}$ is a normal distribution of noise injection with a sample size of $x_i$.
By increasing the SINAD in Eq.~\eqref{eqn:s_snr}, we can observe the inference accuracy of the Neural-PIM accelerator as SINAD changes for different DNNs.
In fact, this noise injection model can be generally applied to other RRAM-based PIM accelerators, e.g., ISAAC~\cite{isaac} and CASCADE~\cite{cascade}.
Although they employ different analog dataflows, the lumped noise can still be modeled as a Gaussian distribution with different variations.

\begin{figure}[!t]
    \centering
    \includegraphics[width=1\linewidth]{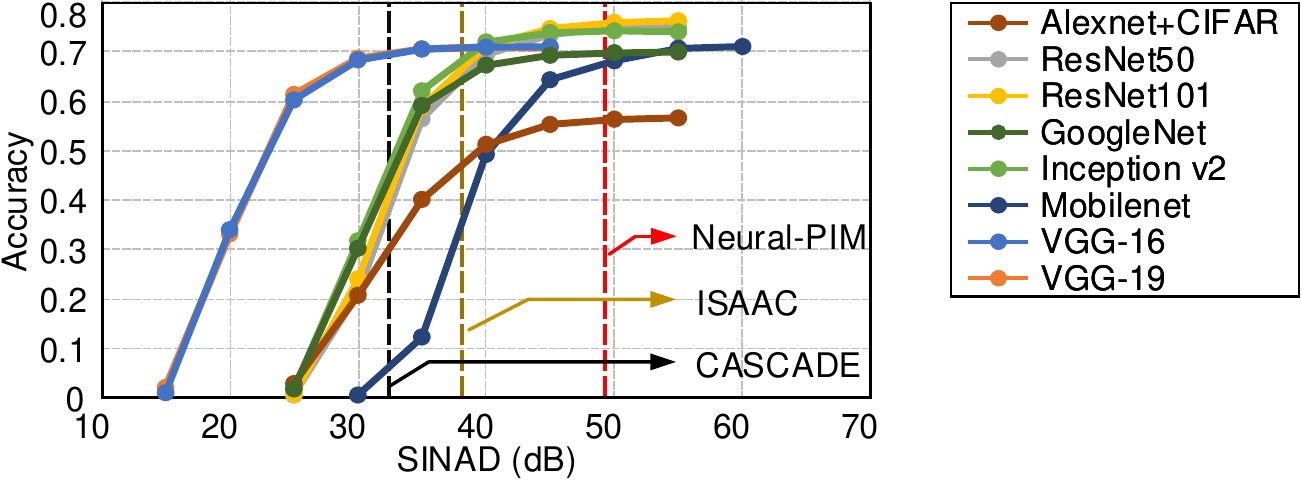}
    \caption{The effect of noise on inference accuracy (top-1 \% on ImageNet). Gaussian noise is added to the activations of NN layers to mimic the affect of hardware noise.} 
    \label{fig:noise_injection}
\end{figure}

Fig.~\ref{fig:noise_injection} demonstrates the effects of varying the SINAD on the inference accuracy of different DNNs with the software-level sweeping. 
Typically, a 45 dB SINAD is the minimum for all DNN models to reach the ideal inference accuracy.
We then simulate the practical SINADs of the analog dataflows employed by the two baseline accelerators, i.e., ISAAC~\cite{isaac} and CASCADE~\cite{cascade}, with the same circuit-level simulation method introduced in Section~\ref{sec: noise_char}.
We mark all SINADs of different analog dataflows in Fig.~\ref{fig:noise_injection} with the vertical lines in different colors. 
The dataflow of CASCADE~\cite{cascade} has the lowest SINAD because it uses 6-bit RRAM cells to buffer analog partial sums.
However, high-precision RRAM cells are particularly vulnerable to the stochastic variation of devices.
ISAAC~\cite{isaac} needs multiple A/D conversions, incurring multiplicative quantization noises due to the repeated A/D conversions within multiple input cycles.
Its dataflow achieves a medium SINAD because the quantization noise is smaller than the noise of a high-precision RRAM cell.
Our dataflow can achieve the highest SINAD with a number of circuit-level optimization techniques as introduced before.
The comparison suggests that the our Neural-PIM accelerator is robust and can guarantee adequate hardware-level inference accuracy with a 50 dB SINAD.

\section{Simulation Methodology}
\label{sec:ex_metho}

In this section, we introduce the simulation methodology used to evaluate the Neural-PIM accelerator.
We first show the baselines for comparison with our proposed accelerator.
We then present the main benchmarks for system evaluations.
Finally, we give the system-level evaluation metrics.

\subsection{Reference Accelerators and Benchmarks}
\label{sec:baseline}
Two baseline accelerators by referring to ISAAC~\cite{isaac} and CASCADE~\cite{cascade} are built and scaled to support 8-bit inference for fair comparison with Neural-PIM accelerator.
All accelerators adopt bit-serial input streaming manner and binary weight mapping method.
Except for the buffer array in the CASCADE-style baseline and the NeuralPeriph circuits in our accelerator, all RRAM devices assume 1-bit precision.

We use 9 well-known DNN benchmarks for evaluations, including 8 CNNs and 1 RNN.
AlexNet~\cite{alexnet}, ResNet-50/ResNet-101~\cite{resnet}, VGG-16/VGG-19~\cite{vgg16}, Inception~\cite{inception}, GoogleNet, and NeuralTalk are fairly selected from previous work~\cite{isaac, cascade}.
We also add a new benchmark, i.e., MobileNet~\cite{mobilenetv2} used by edge devices for evaluation.
All DNN models are trained on ImageNet~\cite{alexnet} with 8-bit quantization.

\begin{figure*}[!t]
    \centering
    \includegraphics[width=1.0\linewidth]{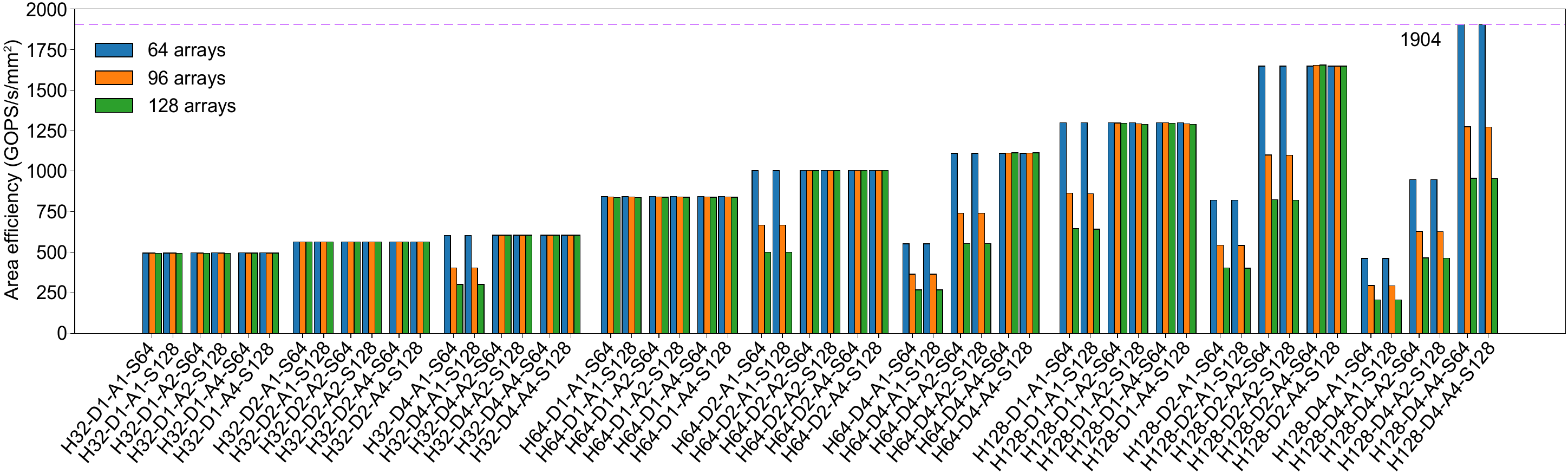}
    \caption{Computation efficiency across the design space exploration. Here, N32-D1-A1-S64 M96 represents that each PE has 96 $32 \times 32$ RRAM crossbars with 1-bit DACs, and one shared NNADC and 64 shared NNS+As for all arrays.} 
    \label{fig:design_space}
\end{figure*}

\subsection{System Simulation}
\label{sec:ea_model}
We evaluate the performances, e.g., accuracy, energy, and throughput of three architectures based on 32 nm technology similar to the previous work~\cite{isaac}.
For two baseline accelerators, we take component models, such as DAC, ADC, S/H, SRAM, and eDRAM buffer from ISAAC~\cite{isaac} and CASCADE~\cite{cascade} and scale the reported performances according to the size and resolution.
For Neural-PIM accelerator, we offline train NN models of NeuralPeriph circuits via stochastic gradient descent with Adam optimizer~\cite{adam} using TensorFlow~\cite{tf}.
During the training, we set the weight precision $A_{\text{R}}$ of NN models be 3-bit and the stochastic variation be $\sigma=0.025$ to guarantee the robustness of NeuralPeriph as proposed by previous works~\cite{pipelineadc}.
Since NNS+A has fixed numbers of inputs and outputs, we sweep the number of hidden neurons to find the optimal structure configurations of NNS+A.
In particular, we find that $H_{\text{S+A}}=12$ can ensure the accuracy and robustness.
Beyond this number, the area overhead of NNS+A will be increased.
After training, the NN models are instantiated on the hardware substrate with 3-bit RRAM devices and CMOS inverters.
The NNADC is taken from our prior work~\cite{pipelineadc}.
The detailed circuit-level simulations of the proposed NeuralPeriph circuits, e.g., accuracy, power, speed, area, and variation are done using 130 nm CMOS in Cadence Spectre.
The performance is then conservatively scaled to 32 nm.
Other components used in Neural-PIM come from ISAAC~\cite{isaac} with proper scaling.

We build a full-system simulator to get energy consumption and throughput of each benchmark.
To evaluate the performance of the accelerators, we use three metrics: energy efficiency ($E$), computation efficiency ($A$) and throughput ($T$). 
Energy efficiency is the number of fixed-point number operations computed per Watt per second (\text{GOPS}/\text{s}/\text{W}).
Computation efficiency is the number of fixed-point number operations computed per $\text{mm}^2$ per second (\text{GOPS}/\text{s}/$\text{mm}^2$).
Throughput is the number of fixed-point number operations computed by per second (GOPS/s).

\section{System Evaluation}
\label{sec:experiment}

In this section, we comprehensively evaluate the proposed Neural-PIM accelerator.
We first perform design space exploration to show the capabilities of the Neural-PIM architecture.
We then perform system-level evaluations with comparisons against the baselines.

\subsection{Design Space Evaluation}
\label{sec:exh}

There are five main hyper-parameters that dominate the performance of our accelerator: 1) the size of RRAM crossbar array $N\times N$, denoted as $N$; 2) the number of crossbar arrays in one PE, denoted as $M$; 3) the number of NNADCs shared by one PE, termed as $A$; 4) the number of NNS+As shared by one crossbar array, termed as $S$; 5) the resolution of DACs, termed as $D$.
The size of eDRAM buffer and the c-mesh width are set correspondingly to limit the search space.
Many of the other parameters, e.g., the width of the bus connecting the eDRAM and the PEs, are derived from the above parameters to maintain correctness
and avoid structural hazards for the worst-case layers. 
The reported peak computation efficiency $A$ in the following assumes that all PEs can be somehow utilized in every cycle.

Fig.~\ref{fig:design_space} shows the computation efficiency under different configurations.
Using larger arrays can improve the computation efficiency since more computations and accumulations are performed.
However, as the number of arrays increase, NeuralPeriph circuits, S/H circuits, and registers, are also raised.
More importantly, the I/O bandwidth limits the number of RRAM arrays.
The peak computation efficiency 1904.0 \text{GOPS}/\text{s}/$\text{mm}^2$ of Neural-PIM accelerator is achieved by 64 $128\times 128$ RRAM arrays in one PE which shares 64 NNS+As and 4 NNADCs with 4-bit DACs for each array.
\textcolor{black}{Table~\ref{tab:parameters} lists the configurations of one tile.}

\begin{table}
    \footnotesize
	\caption{Neural-PIM parameters at the tile level.}

    \label{tab:parameters}
   	\begin{tabular}{c|c|c|c|c}\toprule
   		Comp & Para & Spec & Power ($W$) & Area ($\text{mm}^2$)\\\midrule
        \multicolumn{5}{c}{PE properties (4 PEs per tile)}\\\hline
        NNADC & \begin{tabular}{c} Reso \\ $f$ \\ Num\end{tabular} & \begin{tabular}{c} 8-bit \\ 1.2 GS/s \\ 4\end{tabular} & 6.0$\times$$ 10^{-3}$ & 4.8$\times$$ 10^{-3}$ \\ \hline
        DAC & \begin{tabular}{c} Reso \\ Num\end{tabular} & \begin{tabular}{c} 4-bit\\ 128$\times$64\end{tabular} & 1.0$\times$$10^{-1}$ & 4.3$\times$$10^{-3}$\\ \hline
        S+H & Num & 64$\times$144 & 6.4$\times$$10^{-5}$ & 3.2$\times$$10^{-4}$ \\ \hline
        NNS+A & \begin{tabular}{c} Num\\ $f$\end{tabular} & \begin{tabular}{c}64\\80 MHz \end{tabular} & 1.9$\times$$10^{-2}$ & 4.4$\times$$10^{-2}$ \\ \hline
        Crossbar & \begin{tabular}{c} size \\ Num \end{tabular} & \begin{tabular}{c} 128$\times$128\\ 64\end{tabular} & 9.6$\times$$10^{-2}$ & 1.6$\times$$10^{-3}$ \\ \hline
        IR & Num & 1 & 4.0$\times$$10^{-2}$ & 2.4$\times$$10^{-2}$ \\ \midrule\midrule
    	1 PE & - & - & 1.8$\times$$10^{-1}$ & 8.4$\times$$10^{-2}$\\ \hline
        280 Tiles &-&-& 57.3 & 63.5\\\hline
        Hyper Tr &-&-& 10.4 & 22.88 \\\midrule\midrule
        Total &-&-& 67.7 & 86.4\\\bottomrule
   	\end{tabular}

\end{table}

\subsection{System-Level Performance}
\label{sec:sys_performance}

We use one Neural-PIM chip with 280 tiles to evaluate its energy and throughput.
Each tile consists of 4 PEs and each PE uses the optimal configuration obtained in Section~\ref{sec:exh}.
For a fair comparison with the baselines, all three architectures have the same area.

\begin{figure}[!t]
    \centering
    \includegraphics[width=1\linewidth]{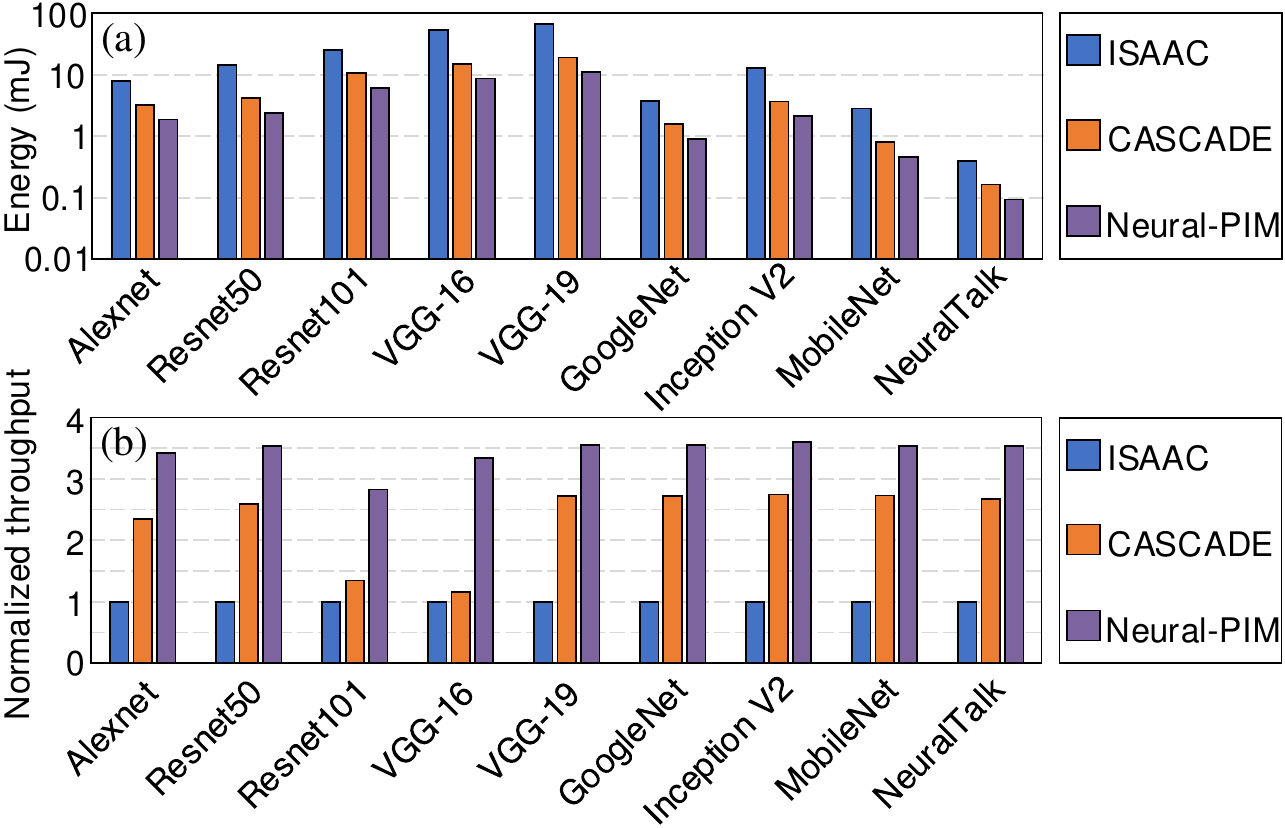}
    \caption{(a), Bench-marked energy consumption and (b), normalized throughput of three accelerators on 8 CNN and one RNN benchmarks.} 
    \label{fig:ener_throu}
\end{figure}

\begin{figure}[!t]
    \centering
    \includegraphics[width=1\linewidth]{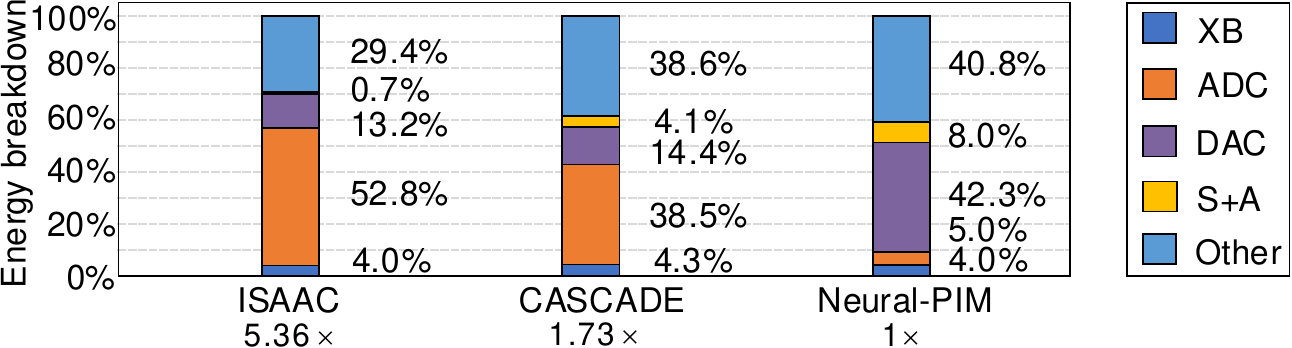}
    \caption{Energy breakdown of the three accelerators.} 
    \label{fig:dnn_breakdown}
\end{figure}

\noindent \textbf{{Energy efficiency.}}
Fig.~\ref{fig:ener_throu}(a) shows the energy consumption of Neural-PIM and the two baselines on the 8 CNN and 1 RNN benchmarks.
The comparison presents the competitive advantage of Neural-PIM: it achieves an average energy efficiency improvement of $1.73\times$ over CASCADE-based architecture~\cite{cascade}, and $5.36\times$ over ISAAC-based architecture~\cite{isaac}.
This improvement is attributed to the fact that Neural-PIM employs the proposed fully analog accumulation scheme, which minimizes the required energy-expensive A/D conversions.
Fig.~\ref{fig:dnn_breakdown} exhibits the energy breakdown of the three accelerators.
The analog summation (``S+A") in Neural-PIM consumes 33$\times$ energy less than the ADCs of ISAAC~\cite{isaac}.

\noindent \textbf{{Throughput.}}
Fig.~\ref{fig:ener_throu}(b) shows the throughput of Neural-PIM and the comparisons with the two baselines on the same DNN benchmarks.
In average, the Neural-PIM architecture achieves $3.43\times$
throughput of the ISAAC-based architecture, and $1.59\times$ throughput of the CASCADE-based architecture.
The improvement comes from the two important features of Neural-PIM.
First, Neural-PIM can achieve faster processing speed with high-resolution DACs for input streaming.
As characterized in Section~\ref{sec:framework}, high-resolution DACs can decrease computation cycles, reducing the processing delay of dot-product.
Second, Neural-PIM can compensate for the area overheads caused by the high-resolution DACs with area-efficient NeuralPeriph circuits.
As elaborated in the following section, Neural-PIM requires much less ADCs at the PE level and all NeuralPeriph circuits are synthesized with RRAM-based hardware substrate.
In this way, Neural-PIM enjoys the higher throughput brought by the high-resolution DACs without sacrificing the density of VMM computation arrays at the PE level.

\noindent \textbf{{Area efficiency.}}
Table~\ref{tb:com_2} lists the configurations of the three architectures at the PE level for comparison.
Although prior arts~\cite{isaac,cascade} do not explicitly report the area efficiency of computing arrays, we include such a metric to better illustrate the different design tradeoff in three architectures.
Specifically, we use the density, i.e., the ratio between the area of all computing arrays in one PE and the total area of one PE as a proxy to measure the area overhead of the peripheral circuits.
ISAAC-based architecture demands an 8-bit ADC for each VMM computing array and uses digital logic circuits for accumulation.
Its VMM computing arrays thus occupies $0.68\%$ of the total PE area.
CASCADE-based architecture requires 3 shared ADCs for all computing arrays in one PE.
In addition, it allocates 4 buffer arrays for each computing array and uses an analog summing-amplifier per buffer array.
Its density is increased to $0.76\%$, because much less ADCs are used in this architecture.
Neural-PIM needs 4 shared NNADCs at the PE level and allocate one NNS+A for each crossbar array.
Its density is $0.71\%$.
Note that we use the area overheads of higher-resolution DACs to trade off the throughput improvement of Neural-PIM.
However, our NeuralPeriph circuits are synthesized by RRAM-based hardware substrate with higher area efficiency than conventional analog circuits, thereby compensating for the area overheads of DACs.
The density of the VMM computing arrays is still comparable to the baselines.
This result also shows that Neural-PIM can improve the system-level throughput with high-resolution DACs but does not incur significant area overheads.

\begin{table}[!tb]
\footnotesize
\centering
\caption{Comparisons of the three architectures at the PE level.}
\begin{tabular}{c|c|c|c}
\toprule
\begin{tabular}[c]{@{}c@{}}Accelerator\\  types\end{tabular}       & \begin{tabular}[c]{@{}c@{}}ISAAC\\ style~\cite{isaac}\end{tabular}           & \begin{tabular}[c]{@{}c@{}}CASCADE\\ style~\cite{cascade}\end{tabular}                     & \begin{tabular}[c]{@{}c@{}}Neural-PIM\\ accelerator \end{tabular}                                                           \\ \midrule
\begin{tabular}[c]{@{}c@{}}Computing \\ configuration\end{tabular} & \multicolumn{3}{c}{\begin{tabular}[c]{@{}c@{}}Inputs: 8-bit; ~Weights: 8-bit;
\\ RRAM computing array: 128$\times$128 with 1-bit cell
\end{tabular}}              \\ \hline
\begin{tabular}[c]{@{}c@{}}Accumulation \\ type\end{tabular}       & Digital                                                          & \begin{tabular}[c]{@{}c@{}}Partially\\ analog\end{tabular} & Analog                                                          \\ \hline
\begin{tabular}[c]{@{}c@{}}Accumulate\\ interface\end{tabular}     & S+A    & \begin{tabular}[c]{@{}c@{}} S+A and\\ Buffer array\end{tabular}        &\begin{tabular}[c]{@{}c@{}} NNS+A\\ \end{tabular}                                                             \\ \hline
\begin{tabular}[c]{@{}c@{}}D/A\\ resolution\end{tabular}     & 1-bit    & 1-bit & 4-bit                                                             \\ \hline
\begin{tabular}[c]{@{}c@{}}A/D \\ resolution\end{tabular}                                                   &  7-bit                                                           & 10-bit                                                                        & 8-bit                                                            \\ \hline
\begin{tabular}[c]{@{}c@{}}Number of\\  ADCs\end{tabular}                                                   & \begin{tabular}[c]{@{}c@{}}64 ADCs \\ per 64 arrays\end{tabular}  & \begin{tabular}[c]{@{}c@{}}3 ADCs \\ per 64 arrays\end{tabular}             & \begin{tabular}[c]{@{}c@{}}4 NNADCs \\ per 64 arrays\end{tabular} \\ \hline
 \begin{tabular}[c]{@{}c@{}} Density  \\ (\# cells/${\text{mm}^2}$) \end{tabular} &  \begin{tabular}[c]{@{}c@{}} 0.68\%\\ (4.5$\times10^6$) \end{tabular}  & \begin{tabular}[c]{@{}c@{}} 0.76\% \\ (5.0$\times10^6$) \end{tabular}  &  \begin{tabular}[c]{@{}c@{}} 0.71\% \\ (4.6$\times10^6$) \end{tabular} \\\bottomrule
\end{tabular}
\label{tb:com_2}
\end{table}

In summary, Neural-PIM can improve both energy efficiency and throughout upon the existing RRAM-based PIM architectures without losing inference accuracy.
Such an architecture adopts an extended analog dataflow to minimize the required A/D conversions using RRAM-based neural approximated peripherals with high area efficiency.

\section{Conclusion}
\label{sec:con}

In this paper, we propose Neural-PIM--a novel architecture for RRAM-based PIM acceleration.
The proposed accelerator adopts an analog dataflow enabled by RRAM-based neural approximated peripherals to minimize the demands of A/D conversions, significantly improving the energy efficiency without hurting area efficiency and throughput.
Simulations using various DNN benchmarks demonstrate that Neural-PIM can achieve an improvement of $5.36\times$ ($1.73\times$) and $3.43\times$ ($1.59\times$) in energy efficiency and in computational throughput respectively, without accuracy loss, as compared to the baseline ISAAC~\cite{isaac} (CASCADE~\cite{cascade}) accelerator with traditional peripherals and digital accumulation schemes.

\section*{Acknowledgment}
This work was partially supported by NSF CCF-1942900, NSF CNS-1739643, National Key Reserch and Development Program of China under Grant 2018YFB1403400, National Natural Science Foundation of China under Grant No. 61834006, and Shanghai Science and Technology Committee under Grant No.18ZR142140. 
W. Cao and Y. Zhao contributed equally to this work.
We thank all reviewers for their constructive comments to improve our work.


\footnotesize
\bibliographystyle{IEEEtran}
\bibliography{main} 






\end{document}